\pdfoutput=1
\documentclass[usenatbib]{mn2e}
\usepackage{hyperref}
\usepackage{apjfonts}
\usepackage{amssymb}
\usepackage{amsmath}
\usepackage{ctable}
\usepackage{url}
\usepackage{breakurl}
\usepackage{fixltx2e} 

\newcommand{\be}{\begin{equation}}
\newcommand{\ee}{\end{equation}}

\newcommand{\gizmourl}{\burl{http://www.tapir.caltech.edu/~phopkins/Site/GIZMO.html}}

\newcommand{\vspacerpostplot}{\vspace{-0.4cm}}
\newcommand{\grainsize}{a_{d}}
\newcommand{\sizeparam}{\alpha}
\newcommand{\Lbox}{L_{\rm box}}

\newcommand\plotonesize[2]
 {\centering \leavevmode \includegraphics[width={#2\columnwidth}]{#1}}
\newcommand{\plotsidesize}[2]
 {\centering \leavevmode \includegraphics[width={#2\textwidth}]{#1}}
\newcommand{\acknowledgments}{\begin{small}\section*{Acknowledgments}\end{small}}
\newcommand\altaffilmark[1]{$^{#1}$}
\newcommand\altaffiltext[1]{$^{#1}$}
\title[Gas-Dust Dynamics in GMCs]{The Fundamentally Different Dynamics of Dust and Gas in Molecular Clouds\vspace{-0.5cm}}

\author[Hopkins \&\ Lee]{
\parbox[t]{\textwidth}{ 
Philip F. Hopkins\altaffilmark{1}\thanks{E-mail:phopkins@caltech.edu} \&\ Hyunseok Lee\altaffilmark{1}
} 
\vspace*{6pt} \\
\altaffiltext{1}{TAPIR, Mailcode 350-17, California Institute of Technology, Pasadena, CA 91125, USA\vspace{-1.1cm}} \\
}

\date{Submitted to MNRAS, October, 2015\vspace{-0.6cm}}
\begin{document}
\maketitle
\label{firstpage}

\begin{abstract}

We study the behavior of large dust grains in turbulent molecular clouds (MCs). In primarily neutral regions, dust grains move as aerodynamic particles, not necessarily with the gas. We therefore directly simulate, for the first time, the behavior of aerodynamic grains in highly supersonic, magnetohydrodynamic turbulence typical of MCs. We show that, under these conditions, grains with sizes $a \gtrsim 0.01\,$micron exhibit dramatic (exceeding factor $\sim 1000$) fluctuations in the local dust-to-gas ratio (implying large small-scale variations in abundances, dust cooling rates, and dynamics). The dust can form highly filamentary structures (which would be observed in both dust emission and extinction), which can be much thinner than the characteristic width of gas filaments. Sometimes, the dust and gas filaments are not even in the same location. The ``clumping factor'' $\langle n_{\rm dust}^{2} \rangle / \langle n_{\rm dust} \rangle^{2}$ of the dust (critical for dust growth/coagulation/shattering) can reach $\sim 100$, for grains in the ideal size range. The dust clustering is maximized around scales $\sim 0.2\,{\rm pc}\,(a/\mu{\rm m})\,(n_{\rm gas}/100\,{\rm cm^{-3}})^{-1}$, and is ``averaged out'' on larger scales. However, because the density varies widely in supersonic turbulence, the dynamic range of scales (and interesting grain sizes) for these fluctuations is much broader than in the subsonic case. Our results are applicable to MCs of essentially all sizes and densities, but we note how Lorentz forces and other physics (neglected here) may change them in some regimes. We discuss the potentially dramatic consequences for star formation, dust growth and destruction, and dust-based observations of MCs.

\end{abstract}

\begin{keywords}
galaxies: formation --- 
star formation: general --- cosmology: theory --- planets and satellites: formation --- accretion, accretion disks --- instabilities --- turbulence
\vspace{-1.0cm}
\end{keywords}

\vspace{-1.1cm}
\section{Introduction}
\label{sec:intro}

Dust is ubiquitous in astrophysics, and critical to understanding phenomena as diverse as star and planet formation, feedback from stars in galaxy formation, and the origin and fate of certain heavy elements. Even if it is only to correct foreground contamination or extinction, understanding the dust size distribution and dust-to-gas ratio, and any possible variations (hence variations in the extinction curve, for example) is necessary to almost every area of astronomy.  

Despite this, there has been little theoretical work to understand the {\em dynamics} of dust as aerodynamic particles in the cold interstellar medium (ISM). For example, a critical process, which could produce fundamentally new phenomena, is the inevitable fluctuation of large dust grain densities in a turbulent medium (so-called ``turbulent concentration''). It is well known that in a primarily neutral, dense gas, massive dust grains (which contain a large fraction of all the ISM metals) behave as aerodynamic particles (the dominant force is drag from collisions with atoms/molecules). As such, they can, under the right conditions, de-couple from the gas, and can clump or disperse independent from gas density fluctuations. 

Much attention has, in fact, been paid to the question of grain density fluctuations (arising from this mechanism and others) in proto-planetary disks. When stirred by turbulence or trapped in various instabilities, the number density of grains can fluctuate by orders of magnitude relative to the gas. This has been seen now in a wide variety of situations, including or excluding grain collisions, in magnetized and non-magnetized disks, and in turbulence driven by self-exciting (``streaming'') instabilities, gravitational instabilities, the magneto-rotational instability, convection, and Kelvin-Helmholtz instabilities \citep[see e.g.][]{bracco:1999.keplerian.largescale.grain.density.sims,cuzzi:2001.grain.concentration.chondrules,youdin.goodman:2005.streaming.instability.derivation,johansen:2007.streaming.instab.sims,carballido:2008.grain.streaming.instab.sims,bai:2010.grain.streaming.sims.test,bai:2010.grain.streaming.vs.diskparams,pan:2011.grain.clustering.midstokes.sims,dittrich:2013.grain.clustering.mri.disk.sims,jalali:2013.streaming.instability.largescales,hopkins:2014.pebble.pile.formation}. In the terrestrial turbulence literature as well, ``preferential concentration'' of aerodynamic particles is well-studied with both laboratory experiments \citep{squires:1991.grain.concentration.experiments,fessler:1994.grain.concentration.experiments,rouson:2001.grain.concentration.experiment,gualtieri:2009.anisotropic.grain.clustering.experiments,monchaux:2010.grain.concentration.experiments.voronoi} and numerical experiments \citep{cuzzi:2001.grain.concentration.chondrules,yoshimoto:2007.grain.clustering.selfsimilar.inertial.range,hogan:2007.grain.clustering.cascade.model,bec:2009.caustics.intermittency.key.to.largegrain.clustering,pan:2011.grain.clustering.midstokes.sims,monchaux:2012.grain.concentration.experiment.review} demonstrating that gas is unstable to the growth of large-amplitude inhomogeneities in the grain density. 

Many authors have pointed out that the relevant phenomena appear to be scale-free, if the dust grains are sufficiently large so that their ``stopping'' (friction or drag) timescale $t_{s}$ corresponds to an eddy turnover time $t_{e}$ for eddies which lie within the inertial range of turbulence \citep{cuzzi:2001.grain.concentration.chondrules,hogan:2007.grain.clustering.cascade.model,bec:2009.caustics.intermittency.key.to.largegrain.clustering,olla:2010.grain.preferential.concentration.randomfield.notes,hopkins:2013.grain.clustering}. Qualitatively, if $t_{s}\ll t_{e}$, grains are well-coupled to gas, so should move with the flow (although they may still ``settle'' and exhibit non-trivial dynamics); if $t_{s} \gg t_{e}$, grains are effectively de-coupled from the local gas flow; when $t_{s}\sim t_{e}$, grains can be ``flung out'' of regions of high vorticity by centrifugal forces, and collect in regions of high strain \citep{yoshimoto:2007.grain.clustering.selfsimilar.inertial.range,bec:2008.markovian.grain.clustering.model,wilkinson:2010.randomfield.correlation.grains.weak,gustavsson:2012.grain.clustering.randomflow.lowstokes}. 

In principle, the same mechanisms which would generate large grain-density fluctuations in a protoplanetary disk with $\sim1-100\,$cm boulders could operate on micron-sized dust in a giant molecular cloud (GMC). In fact, \citet{hopkins:totally.metal.stars} applied the analytic scalings derived as a function of the dimensionless ratio of grain stopping time to dynamical time in proto-planetary disks to GMCs, and argued that this implied micron-sized dust could cluster strongly on scales large enough to alter stellar abundances (potentially dramatically, in the most extreme cases). 

However, essentially all of of the numerical and experimental studies to date have focused on sub-sonic, either incompressible or weakly-compressible gas with inefficient cooling (adiabatic), usually without magnetic fields. In the cold ISM, on the other hand, the gas is rapidly cooling (effectively close to isothermal), magnetized, highly compressible (with density fluctuations of factors of thousands), and super-sonically turbulent with Mach numbers $\gtrsim 10$ on the scales of large clouds. Moreover, in incompressible gas, the velocity field is divergence-free, so the vorticity and strain dominate grain aggregation; in highly super-sonic turbulence, the density field is a network of filamentary shocks and rarefactions, each of which can trap or disperse dust \citep{booth:dust.gas.2d.shock.tests}. And the ``stopping time'' $t_{s}$ in the supersonic case is not a constant (as it is for grains of a given size in a sub-sonic medium), but depends on the local density and dust-gas relative velocity. It is not obvious that the dynamics should be even qualitatively similar in these cases.

In this paper, we therefore for the first time explore the dynamics of dust grains in a neutral, highly-compressible, supersonically turbulent medium, under conditions which resemble observed atomic/molecular clouds, clumps, and cores. We show that a wide range of dust grain sizes show dramatic clustering effects, in many cases more than would be expected from simply scaling up the sub-sonic case, and discuss the implications for different observations and theoretical models.

\begin{figure*}
    \plotsidesize{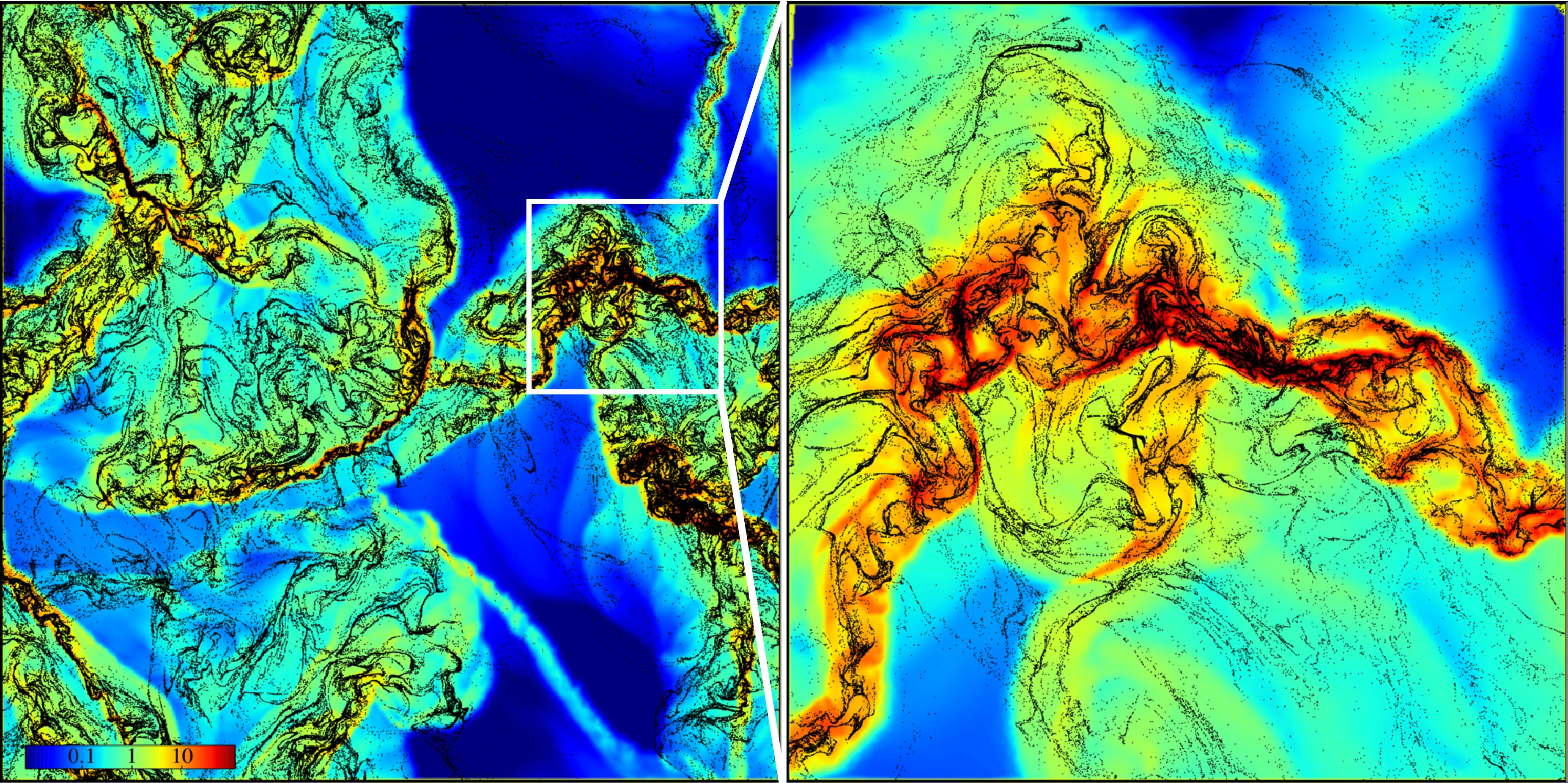}{0.95}
    \vspace{-0.25cm}
    \caption{Image of a very high-resolution ($1024^{2}$) 2D simulation of aerodynamic grains in supersonic MHD turbulence, here with grain size parameter $\sizeparam \equiv (\bar{\rho}_{d}\,\grainsize) / (\langle \rho_{\rm gas} \rangle \,\Lbox) =0.01$ (see Eq.~\ref{eqn:grain.size} for how this relates to physical grain sizes $a\sim0.001-1\,\mu$m)  and rms Mach number $\mathcal{M}\sim5$. Time shown is after the rms Mach numbers and magnetic energy reach steady-state. We show the full simulation box ({\em left}) and zoom-in of a dense region ({\em right}). 
     Color shows gas density relative to the mean ($n_{\rm gas}/\langle n_{\rm gas} \rangle$), on a logarithmic scale (see colorbar); black points show the dust (super)-particles. As expected, gas forms a filamentary network of shocks and rarefactions. Dust loosely traces the same on large scales, but with much more detailed small-scale structure. Much of the dust lies along razor-thin filaments, some of which are not associated with a gas filament.
        \vspacerpostplot 
    \label{fig:demo}}
\end{figure*}

\begin{figure}
    \plotonesize{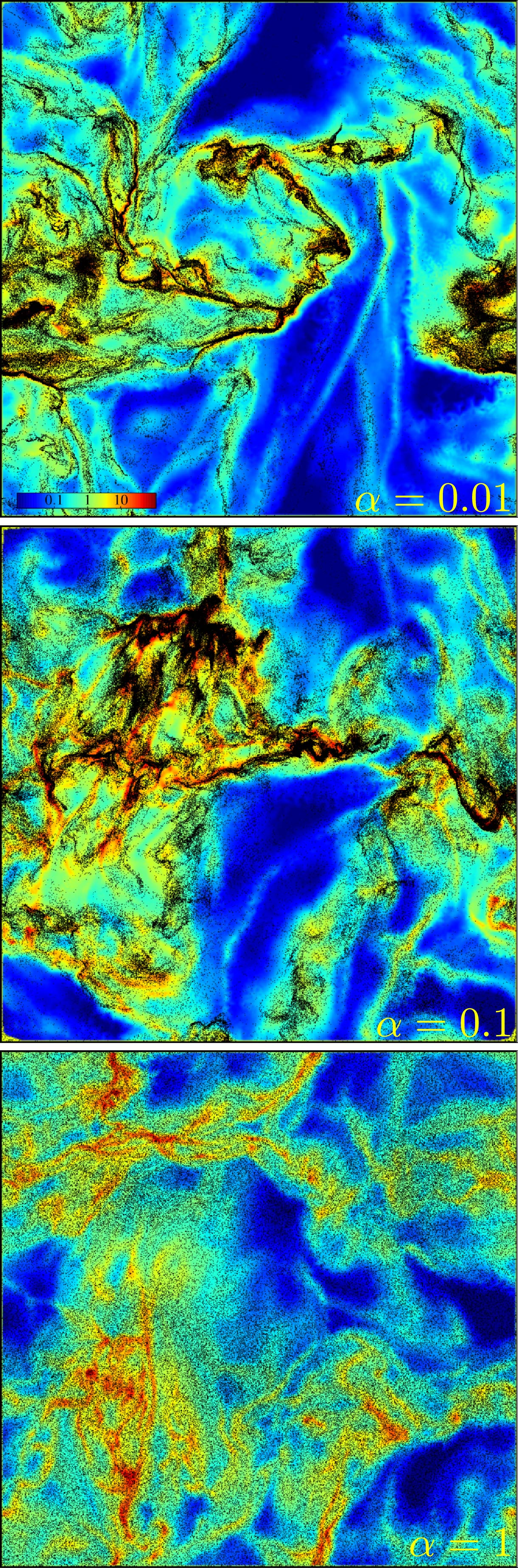}{0.82}
    \vspace{-0.25cm}
    \caption{Images (as Fig.~\ref{fig:demo}) of our standard 3D, $256^{3}$ MHD simulations (each shows a thin slice through $z=0$), here with Mach number $\mathcal{M}\sim10$. Time is after run each reaches steady-state; color shows gas density; black points show dust. Each image shows a different dust size $\alpha=0.01,\,0.1,\,1$ (top-to-bottom). For small $\alpha \lesssim 0.1$, dust traces gas roughly on large scales, but shows very thin, small-scale filaments which can be over-dense relative to the gas (seen in Fig.~\ref{fig:demo}). Intermediate $\alpha\sim0.1$ exhibit more dramatic large-scale dust clumping. Large $\alpha\sim 1$ dust is only weakly coupled to the gas, and remains at approximately the mean density everywhere.
        \vspacerpostplot 
    \label{fig:demo.vs.alpha}}
\end{figure}

\vspace{-0.5cm}
\section{Methods}
\label{sec:methods}

\subsection{Magnetohydrodynamics \&\ Turbulent Driving}

Our simulations use {\small GIZMO} \citep{hopkins:gizmo},\footnote{A public version of this code is available at \gizmourl.}
a mesh-free, Lagrangian finite-volume Godunov code designed to capture advantages of both grid-based and smoothed-particle hydrodynamics (SPH) methods, built on the gravity solver and domain decomposition algorithms of {\small GADGET-3} \citep{springel:gadget}. In \citet{hopkins:gizmo} and \citet{hopkins:mhd.gizmo} we consider extensive surveys of test problems in both hydrodynamics and MHD, and demonstrate accuracy and convergence in good agreement with well-studied regular-mesh finite-volume Godunov methods and moving-mesh codes \citep[e.g.\ {\small ATHENA} \&\ {\small AREPO};][]{stone:2008.athena,springel:arepo}. We run {\small GIZMO} in its Meshless-Finite Mass (MFM) mode but have verified that Meshless Finite-Volume (MFV) mode produces nearly identical results (as expected from the previous studies). Note that in \citet{hopkins:gizmo,hopkins:mhd.gizmo}, we demonstrate excellent agreement between {\small GIZMO} and high-resolution, state-of-the-art moving mesh and grid-based codes for simulations of {\em both} super-sonic and sub-sonic MHD turbulence.

The turbulent driving routines follow \citet{bauer:2011.sph.vs.arepo.shocks}. Briefly, a periodic box is stirred via the usual method in e.g.\ \citet{schmidt:2008.turb.structure.fns,federrath:2008.density.pdf.vs.forcingtype,price:2010.grid.sph.compare.turbulence}, where a small range of modes corresponding to wavelengths between $1/2-1$ times the box size are driven in Fourier space as an Ornstein-Uhlenbeck process, with the compressive part of the acceleration projected out via a Helmholtz decomposition in Fourier space so that the driving is an adjustable mix of compressible modes and incompressible/solenoidal modes. Our specific implementation has been verified for both hydro and MHD cases in \citet{hopkins:lagrangian.pressure.sph,hopkins:gizmo,hopkins:mhd.gizmo}. Our parameter choices for the driving generally follow \citet{bauer:2011.sph.vs.arepo.shocks}, Table~4, but we specify the relevant Mach numbers below. We initialize a uniform seed field ${\bf B} = B_{0}\hat{z}$, which determines the saturated mean-field strength.

\vspace{-0.5cm}
\subsection{Dust Dynamics}

We follow \citet{carballido:2008.grain.streaming.instab.sims,hogan:1999.turb.concentration.sims,johansen:2007.streaming.instab.sims,johansen:2009.particle.clumping.metallicity.dependence,bai:2010.grain.streaming.sims.test,pan:2011.grain.clustering.midstokes.sims} and model the dust via a collection of ``super-particles,'' each one of which represents an ensemble of grains of a fixed size, whose trajectories are integrated on-the-fly through the fluid. This is essentially a Monte Carlo ``tracer particle'' approach. 

\citet{draine.salpeter:ism.dust.dynamics} show that grains obey the following equation of motion: 
\begin{align}
\label{eqn:grain.eom} \frac{d{\bf u}_{d}}{dt} &= -\frac{{\bf u}_{d}-{\bf u}_{\rm gas}}{t_{s}} \\ 
\label{eqn:tstop} t_{s} &\equiv \frac{\pi^{1/2}}{2\sqrt{2}}\,\left( \frac{\bar{\rho}_{d}\,\grainsize}{c_{s}\,\rho_{\rm gas}} \right)\,\left( 1 + \left|\frac{3\pi^{1/2}}{8}\,\frac{{\bf u}_{d}-{\bf u}_{\rm gas}}{c_{s}} \right|^{2}\right)^{-1/2}
\end{align}
where ${\bf u}_{d}$ is the grain velocity, $d/dt$ is a Lagrangian derivative, $c_{s}$ and $\rho_{\rm gas}$ the isothermal sound speed and density of the gas, $\bar{\rho}_{d}\approx2.4\,{\rm g\,cm^{-3}}$ is the internal (material) grain density \citep{draine:2003.dust.review}, and $\grainsize \sim 0.001-1\,\mu{\rm m}$ is the grain radius.\footnote{Eq.~\ref{eqn:grain.eom} assumes grains are in the Stokes, rather than the Epstein, limit for drag, i.e.\ $\grainsize < (9/4)\,\lambda$, where $\lambda=(n_{\rm gas}\,\sigma_{\rm gas})^{-1}$ is the gas mean free path. For molecular hydrogen cross-sections at $\sim30\,$K, this requires $\grainsize < 10^{13}\,{\rm cm}\,(n_{\rm gas}/10\,{\rm cm^{-3}})$, so is obviously satisfied.} 

Note that in the sub-sonic limit, this becomes the well-studied Stokes expression with constant ``stopping time'' $t_{s}$. In the super-sonic case, $c_{s}$ and $\rho_{\rm gas}$ depend on position (since the gas is compressible), and the term in ${\bf u}_{d}-{\bf u}_{\rm gas}$ can be important. This replaces the sound speed with the ``total gas-dust'' velocity $\sim \sqrt{c_{s}^{2} + |{\bf u}_{d} - {\bf u}_{\rm gas}|^{2}}$. 

We solve this equation by kernel-interpolating the quantities $c_{s}$, $\rho_{\rm gas}$, and ${\bf u}_{\rm gas}$ and their derivatives from gas particle positions (where they are determined by the MHD solver) to the grain particle position, i.e.\ $\rho_{\rm gas}({\bf x}_{d,\,i}) = \sum\,W({\bf x}_{{\rm gas},\,j} - {\bf x}_{d,\,i},\,h_{i})\,\rho_{\rm gas}({\bf x}_{{\rm gas},\,j})$ where $W$ is normalized so that $1=\sum\,W({\bf x}_{{\rm gas},\,j} - {\bf x}_{d,\,i},\,h_{i})$. The functional form of $W$ and kernel size $h$ are identical to those used for hydrodynamic operations \citep[see][]{hopkins:gizmo}, so the interpolation is numerically stable and consistent. Eq.~\ref{eqn:grain.eom} is then solved exactly over half-timesteps $\Delta t/2$, assuming the interpolated quantities vary linearly in time and space, and we use this to determine the mean numerical acceleration over the corresponding interval $\bar{{\bf a}}_{i}(t,\,t+\Delta t/2) = [{\bf u}_{d}^{\rm exact}(t+\Delta t/2)-{\bf u}_{d}(t)]/[\Delta t/2]$. This maintains good behavior even in the limit $t_{s}\ll \Delta t$. The particle trajectories are then integrated with a semi-implicit leapfrog scheme (this is already well-tested in our code for collisionless particles in e.g.\ cosmological simulations). To ensure numerical stability, grain particles obey the usual timestep limits for all collisionless particles (e.g.\ $dt < {\rm MIN}[0.1\,(h/|d{\bf u_{d}}/dt|)^{1/2},\,0.2\,|\nabla\cdot{\bf u}_{d}|^{-1}]$), plus a Courant criterion given by the minimum of the timestep of neighbor gas particles or $0.05\,h/\sqrt{c_{s}^{2}+|{\bf u}_{\rm gas}-{\bf u}_{d}|^{2}}$.

\vspace{-0.5cm}
\subsection{Units}
\label{sec:units}

We adopt an isothermal equation of state ($\gamma=1$) for the gas; this is a reasonable assumption for molecular clouds over the density and temperature range of interest here, and enables more direct comparison with previous studies of supersonic turbulence and star formation \citep[see e.g.][]{li:2005.turb.reg.sfr,krumholz:2007.rhd.protostar.modes,federrath:2008.density.pdf.vs.forcingtype,schmidt:2009.isothermal.turb,kritsuk:2011.mhd.turb.comparison,molina:2012.mhd.mach.dispersion.relation,hopkins:2012.intermittent.turb.density.pdfs,konstantin:mach.compressive.relation}. 

With this assumption, the inviscid ideal MHD equations are inherently scale-free. The box length $\Lbox$, mean gas density $\langle \rho_{\rm gas} \rangle \equiv M_{\rm gas} / \Lbox^{3}$, and sound speed $c_{s}$ can therefore be freely rescaled to any physical values. Physically, this means our results are {\em entirely} determined by three dimensionless numbers: the box-averaged Mach number $\mathcal{M} \equiv \langle |{\bf u}_{\rm gas}|^{2} \rangle^{1/2} / c_{s}$, the (coherent) magnetic mean-field strength $|\langle {\bf B} \rangle| / (\rho^{1/2}\,c_{s})$, and the ``grain size parameter'' $\sizeparam$:
\begin{align}
\sizeparam \equiv \frac{\bar{\rho}_{d}\,\grainsize}{\langle \rho_{\rm gas} \rangle \,\Lbox}
\end{align} 
(i.e.\ the value of $\bar{\rho}_{d}\,\grainsize$ in code units; note that only this combination of grain size and density appears in the equations, so $\grainsize$ and $\bar{\rho}_{d}$ are formally degenerate). 

A grain parameter $\sizeparam$ translates to physical grain size $\grainsize$ as:
\begin{align}
\label{eqn:grain.size} \grainsize = 0.4\,\sizeparam\,\mu{\rm m}\,\left( \frac{\Lbox}{10\,{\rm pc}} \right)\,\left( \frac{\langle n_{\rm gas} \rangle}{10\,{\rm cm^{-3}}} \right)\,\left( \frac{\bar{\rho}_{d}}{2.4\,{\rm g\,cm^{-3}}} \right)^{-1}
\end{align}

To aid in rescaling to physical units, if we assume the simulations sample Milky Way-like GMCs that lie on the observed linewidth-size relation ($\mathcal{M} \sim (R/R_{\rm sonic})^{1/2}$ with $R_{\rm sonic} \sim 0.1\,{\rm pc}$ being the sonic length), and size-mass relation ($M_{\rm GMC} \propto R_{\rm GMC}$, or $\langle \Sigma_{\rm GMC} \rangle \sim 300\,M_{\sun}\,{\rm pc^{-2}}\sim$\,constant), and that our boxes sample ``typical'' sub-regions of the clouds, we obtain:
\begin{align}
\Lbox &\sim 10\,{\rm pc}\,\left(\frac{\mathcal{M}_{\rm box}}{10}\right)^{2}\,\left( \frac{R_{\rm sonic}}{0.1\,{\rm pc}} \right) \\ 
\langle n_{\rm gas} \rangle &\equiv \frac{\langle \rho_{\rm gas} \rangle}{\mu\,m_{p}} \sim 10\,{\rm cm^{-3}}\,\left( \frac{\langle \Sigma_{\rm GMC}\rangle }{100\,M_{\odot}\,{\rm pc^{-2}}} \right)\,\left( \frac{R_{\rm GMC}}{100\,{\rm pc}} \right)^{-1} 
\end{align}
where we assume a mean molecular weight $\mu\approx2.3$ to convert between $\rho_{\rm gas}$ and $n_{\rm gas}$.

Moreover, to the extent that turbulence is (approximately) self-similar, we can think of our lower-Mach number simulations as sampling ``sub-volumes'' of our higher-Mach number simulations (similar to how the effective box size scales with Mach number above, if we assume a linewidth-size relation). At infinite resolution, a sufficiently large box of high $\mathcal{M}$ should contain all possible realizations of smaller-$\mathcal{M}$ sub-regions. Moreover, in the limit where the initial mean-field is weak ($|\langle {\bf B} \rangle| / (\rho^{1/2}\,c_{s}) \lesssim 1$), the magnetic field dynamics are dominated by those produced by the turbulent dynamo itself ($\langle |{\bf B}| \rangle \gg |\langle {\bf B} \rangle |$) and are just a function of $\mathcal{M}$, and the mean-field value is irrelevant. This is (by construction) the case in many of our simulations, although we also consider strong mean-field cases and show they have weak effects on grain clustering. 

So in a sense, there is really one dominant dimensionless parameter ($\sizeparam$) which specifies the physics. But because we are limited by computational cost, it is more convenient to run separate boxes of different $\mathcal{M}$; given the linewidth-size relation above, resolving the same smallest physical scale as in one of our $256^{3}$, $\mathcal{M}=2$ boxes in a $\mathcal{M}=10$ box would require a $\sim 6400^{3}$ (trillion-particle) simulation! 

Because we do not explicitly include the ``back-reaction'' of dust grains on gas, the absolute value of the dust-phase metallicity (or equivalently, the mean dust abundance) $Z_{d} \equiv \rho_{\rm dust} / \rho_{\rm gas} = (4\pi/3\,\bar{\rho}_{d}\,\grainsize^{3}\,n_{\rm dust}) / (\mu\,m_{p}\,n_{\rm gas})$ does not enter our equations. The simulations predict relative fluctuations in $Z_{d}$ but can be freely rescaled to any mean $\langle Z_{d} \rangle$, modulo caveats below.

\begin{footnotesize}
\ctable[
  caption={{\normalsize 3D Simulations}\label{tbl:sims}},center]{cccccc}{
\tnote[ ]{Parameters describing the simulations analyzed in the text: Each has unit box length, sound speed, and mean density in code units (see \S~\ref{sec:units} for physical units), and is a 3D MHD simulation with driven isothermal turbulence assuming a natural mix of compressive and solenoidal modes. All runs have $256^{3}$ gas and $2\times256^{3}$ dust particles. \\
{\bf (1)} $\sizeparam$: Grain size parameter (see Eq.~\ref{eqn:grain.size} to relate this to physical grain size). Most runs adopt a single $\sizeparam$; however we include one run adopting a uniform logarithmic (random) distribution of $\alpha$ between $0.005-0.015$, to test the effects of non-uniform sizes. \\
{\bf (2)} $\mathcal{M}$: Steady-state turbulent rms Mach number on the box scale (set by the driving routines). \\
{\bf (3)} $\mathcal{M}_{A}$: Steady-state volume-averaged Alfv{\'e}n Mach number $\mathcal{M}_{A} = \langle |{\bf v}| / v_{A} \rangle$, where the Alfv{\'e}n speed $v_{A} = |{\bf B}|/{\rho^{1/2}}$. \\
{\bf (4)} $| \langle {\bf B} \rangle |$: Time-averaged mean-field strength (in code units): $| \langle {\bf B} \rangle |_{\rm phys} \sim 1.3\,\mu G\,| \langle {\bf B} \rangle |_{\rm code}\,\sqrt{(\langle n_{\rm gas} \rangle/10\,{\rm cm^{-3}})\,(T/100\,{\rm K})}$. Note that rms field strengths can be much larger. \\
{\bf (5)} $C$: Time-averaged dust clumping factor 
$C \equiv \langle n_{\rm dust}^{2} \rangle / \langle n_{\rm dust} \rangle^{2}$. \\
{\bf (6)} $C$ (Dense Gas): Dust clumping factor measured only in the dense ($n_{\rm gas} > \langle n_{\rm gas} \rangle$) gas. \\ 
}
}{
\hline\hline
\multicolumn{1}{c}{Grain} &
\multicolumn{1}{c}{Mach} &
\multicolumn{1}{c}{Alfv{\'e}n} &
\multicolumn{1}{c}{Mean} &
\multicolumn{1}{c}{Clumping} &
\multicolumn{1}{c}{Clumping} \\
\multicolumn{1}{c}{Size} &
\multicolumn{1}{c}{Number} &
\multicolumn{1}{c}{Mach} &
\multicolumn{1}{c}{Field} &
\multicolumn{1}{c}{Factor} &
\multicolumn{1}{c}{Factor} \\
\multicolumn{1}{c}{$\sizeparam$} &
\multicolumn{1}{c}{$\mathcal{M}$} &
\multicolumn{1}{c}{$\mathcal{M}_{A}$} &
\multicolumn{1}{c}{$| \langle {\bf B} \rangle |$} &
\multicolumn{1}{c}{$C$} &
\multicolumn{1}{c}{(Dense Gas)} \\
\hline
0.001 & 10 & 6 & 0.2 & 32 & 42 \\
0.01 & 10 & 6 & 0.2 & 57 & 78 \\ 
0.005-0.015 & 10 & 6 & 0.2 & 49 & 67 \\ 
0.03 & 10 & 6 & 0.2 & 73 & 110 \\
0.10 & 10 & 6 & 0.2 & 39 & 68 \\
1.0 & 10 & 6 & 0.2 & 4.9 & 6.2 \\
\hline
0.01 & 2 & 5 & 0.05 & 21 & 28 \\ 
0.1 & 2 & 5 & 0.05 & 23 & 33 \\ 
1.0 & 2 & 5 & 0.05 & 4.7 & 5.6 \\ 
\hline
0.03 & 10 & $\infty$ &  0 & 52 & 74 \\ 
0.03 & 10 & 6 & 0.5 & 73 & 110 \\
0.03 & 10 & 2.5 & 4.3 & 75 & 120 \\ 
0.03 & 10 & 0.4 & 27 & 55 & 93 \\ 
\hline\hline\\
}
\end{footnotesize}


\begin{figure*}
    \plotsidesize{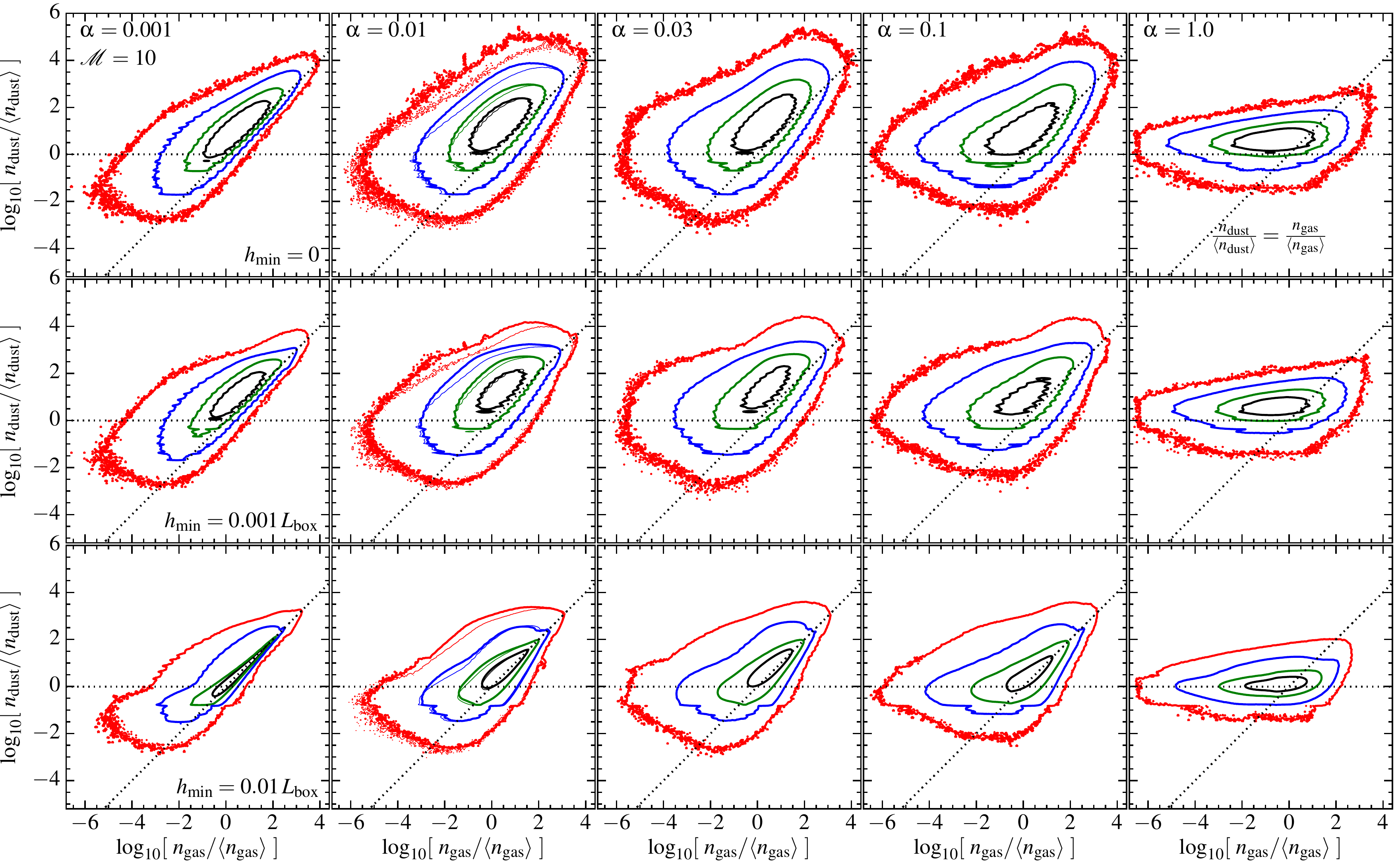}{0.97}
    \vspace{-0.25cm}
    \caption{Distribution of dust and gas densities in the 3D, Mach number $\mathcal{M}\sim10$ (Alfv{\'e}n $\mathcal{M}_{A}\sim6$) simulations from Fig.~\ref{fig:demo.vs.alpha}. We measure dust and gas density around each dust particle, at all times after the system reaches steady-state. We plot iso-density contours at fixed probability density levels $dP/d\log{n_{\rm gas}}\,d\log{n_{\rm dust}} = 10^{-1},\,10^{-2},\,10^{-4},\,10^{-7}$ (black, green, blue, red, respectively). Each column shows a different grain size parameter $\sizeparam=0.001,\,0.01,\,0.03,\,0.1,\,1$ ({\em left-to-right}). Each row smooths the density around each particle on a progressively larger scales corresponding to a spherical radius $h_{\rm min}=0$ (no smoothing, {\em top}), $h_{\rm min}=0.001\,L_{\rm box}$ ({\em middle}), and $h_{\rm min}=0.01\,L_{\rm box}$ (approximately the sonic length $R_{\rm sonic}$; {\em bottom}). We show $n_{\rm dust}=\langle n_{\rm dust} \rangle$ (dust at constant density) and $n_{\rm dust} = (\langle n_{\rm dust} \rangle/\langle n_{\rm gas} \rangle)\,n_{\rm gas}$ (perfect dust-gas coupling, i.e.\ $\delta = 1$) as dotted lines.    
    At high $\alpha\gtrsim1$, the dust is de-coupled from the gas and remains close to $\langle n_{\rm dust} \rangle$ independent of $n_{\rm gas}$. At low $\alpha\ll 0.001$, the dust is tightly coupled to the gas, with significant scatter on very small scales that is quickly averaged-out on larger scales. At intermediate $\alpha\sim0.01-0.1$, the dust de-couples from gas at low gas densities, and traces it on average at high densities, but with large fluctuations in $n_{\rm dust}/n_{\rm gas}$ even at the highest gas densities. These fluctuations are on larger scales for the larger $\alpha$, and for $\alpha\gtrsim 0.01$ are only weakly averaged-down as we smooth over scales as large as $\sim R_{\rm sonic}$. 
    For $\alpha=0.01$, we also compare ({\em thin lines}) one case where the grains have not a single size but a random distribution of sizes between $\alpha=0.005-0.015$; because grains in the same locations experience different drag, this slightly reduces the maximum dust clustering, but the effect is weak.    
    \vspacerpostplot 
    \label{fig:ndust.ngas.m10}}
\end{figure*}

\vspace{-0.5cm}
\subsection{Neglected Dust Physics}

We neglect several processes in this study: 

\begin{itemize} 

\item{Dust-dust collisions: The mean free path to these is $\sim(n_{\rm dust}\,\sigma_{d})^{-1}\sim \sizeparam\,Z_{d}^{-1}\,\Lbox$ (see \S~\ref{sec:units} for definitions); stopping/deceleration lengths in the gas are $\sim \sizeparam\,(1+\mathcal{M}^{2})^{-1/2}\,\Lbox$, so dust-dust collisions are sub-dominant by a factor $\sim Z_{d}/(1+\mathcal{M}^{2})^{1/2} \ll 1$. This does not mean dust collisions are uninteresting, as they can play a key role in modifying the dust size distribution over time. But while the dust dynamics we study may critically alter dust collisions, dust collisions do not (usually) significantly alter the dust dynamics.}

\item{Destruction/creation: We study dust dynamics in cold cloud regions, so we do not consider sources of new dust and/or destruction by shocks/sputtering from SNe and stellar winds, although these can occur inside GMCs later in the cloud lifetime. Turbulent shocks inside the cold cloud regions do not reach sufficient temperatures to destroy grains \citep{draine:dust.destruction}.}

\item{Coulomb forces: Following \citet{draine.salpeter:ism.dust.dynamics}, for low-temperature ($T\lesssim 10^{5}\,K$) gas we expect the ratio of Coulomb forces to collisional drag (for grains in the size range of interest here) to be $\sim 10\,f_{\rm ion}$, where $f_{\rm ion}\lesssim 10^{-7}$ is the ionized fraction of gas in GMCs, so this is negligible.}

\item{Radiation pressure: Near massive stars, this can dominate grain dynamics, but not in random portions of the cloud. Assuming geometric absorption, the ratio of radiation pressure to drag forces at a distance $r_{\ast}$ from an O-star (luminosity $L_{\ast}$) is $\sim 0.5\,(L_{\ast}/10^{4}\,L_{\sun})\,(n_{\rm gas}/10\,{\rm cm^{-3}})^{-1}\,(r_{\ast}/{\rm pc})^{-2}\,(|{\bf u}_{d}-{\bf u}_{\rm gas}|/10\,{\rm km\,s^{-1}})^{-2}$. The radius where radiation pressure dominates is within the Stromgren sphere (i.e.\ fundamentally different conditions from what we simulate) for essentially all stars.}

\item{Lorentz forces: The acceleration from the Lorentz force is $d{\bf u}_{d}/dt = (z_{d}\,e/m_{d}\,c)\,({\bf u}_{d}-{\bf u}_{g})\times{\bf B}$ (where $z_{d}\,e$ is the grain charge, $m_{d}$ its mass, ${\bf B}$ the local magnetic field, and $c$ the speed of light); if we combine this with the expectation value of $\langle z_{d} \rangle$ calculated by \citet{draine:1987.grain.charging} for grains in cold molecular clouds (as a function of grain size and temperature), the ratio of Lorentz to collisional forces becomes $\sim 0.1\,B_{\mu {\rm G}}\,\langle z_{d} \rangle\,a_{\mu{\rm m}}^{-2}\,n_{10\,{\rm cm^{-3}}}^{-1}\,T_{100\,{\rm K}}^{-1/2}\,x^{-1/2} \sim B_{\mu{\rm G}}\,T_{100\,{\rm K}}^{1/2}\,a_{\mu{\rm m}}^{-1}\,n_{10\,{\rm cm^{-3}}}^{-1}\,x^{-1/2}$, where $x\equiv 1+9\pi\,|{\bf u}_{d}-{\bf u}_{g}|^{2}/64\,c_{s}^{2}$. If magnetic fields follow the expected equipartition of the supersonic turbulent dynamo ($E_{B}\sim 0.05\,E_{K}$; see \citealt{kritsuk:2011.mhd.turb.comparison,federrath:supersonic.turb.dynamo}), then we can further simplify and obtain $\sim 0.1\,T_{30\,{\rm K}}\,a_{\mu{\rm m}}^{-1}\,n_{10\,{\rm cm^{-3}}}^{-1/2}$. Therefore, over much of the physical parameter space of interest, Lorentz forces are sub-dominant to collisional drag. However, they are by no means negligible and clearly could be dominant in some regimes \citep[see e.g.][]{yan.2004:lorentz.forces.drag.dust.ism.analytic}. In future work, we will extend our simulations with explicitly coupled Lorentz force equations, but these are complicated to include with drag in a numerically stable manner and depend on some model for grain charging, so we will neglect them for now (with the appropriate caveats). }

\item{Back-Reaction: We neglect the loss of momentum from the gas to the grains, which scales with the dust-to-gas mass ratio $Z_{d}$. Unlike the proto-planetary disk case, where grain concentrations might reach $Z_{d}\gg100$, we have $\langle Z_{d} \rangle \sim0.01 \ll 1$; so this is usually negligible. It is always negligible for sufficiently low-metallicity clouds. However the maximum dust concentrations we simulate do correspond to $Z_{d}\gg1$ if the cloud has solar metallicity, so in future work we will also consider this in more detail.}

\end{itemize}

\begin{figure}
    \plotonesize{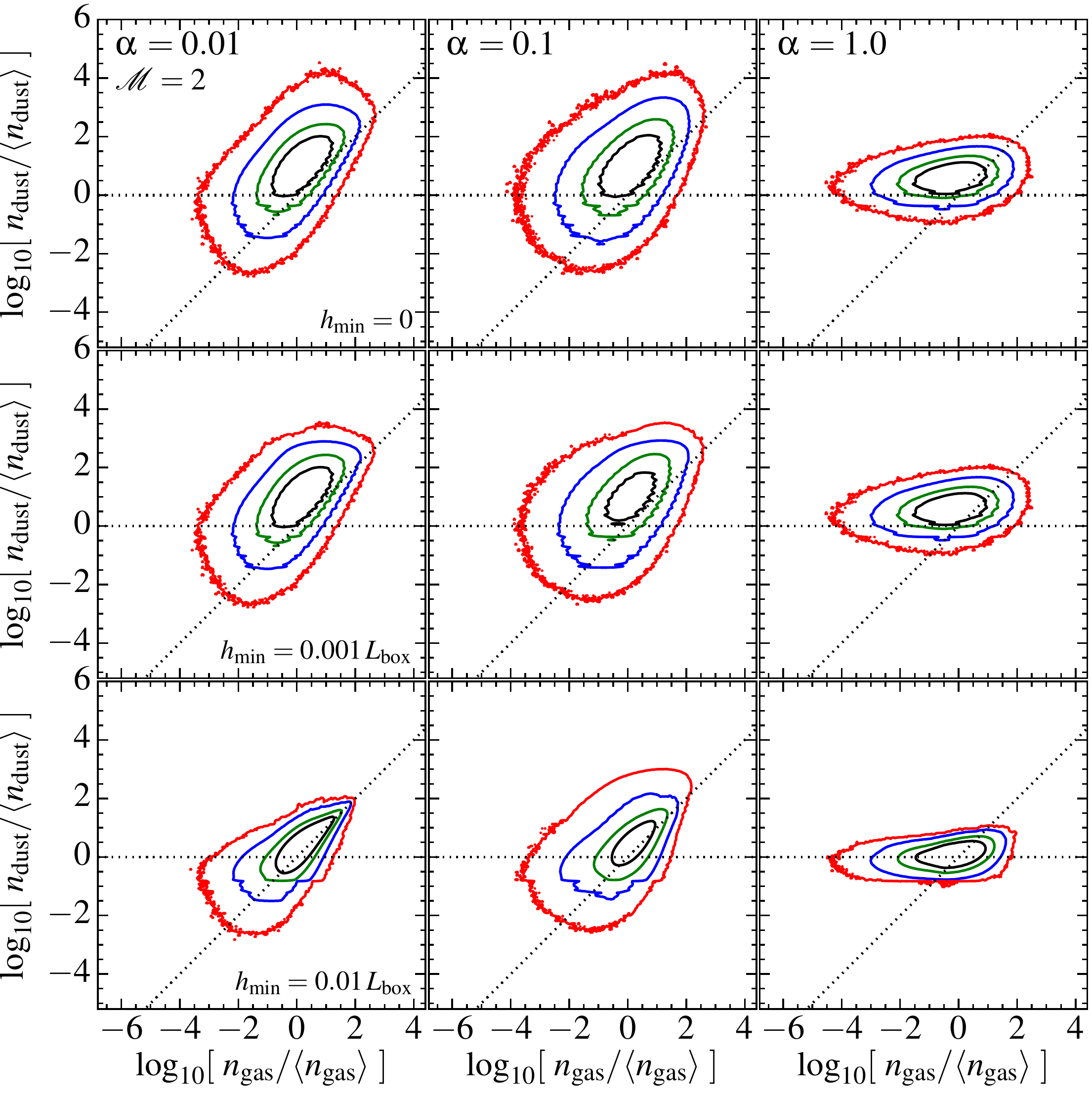}{0.97}
    \vspace{-0.25cm}
    \caption{Distribution of dust and gas densities in the 3D, Mach number $\mathcal{M}\sim2$ simulations; style is identical to Fig.~\ref{fig:ndust.ngas.m10}. As expected, the lower Mach numbers produce much smaller gas-density fluctuations, so the dynamic range sampled is significantly smaller than Fig.~\ref{fig:ndust.ngas.m10}. At fixed $n_{\rm gas}$ within the range probed, the dynamics are similar, and the dust-to-gas ratio fluctuates by a similar (albeit slightly smaller) amount, compared to $\mathcal{M}=10$.
    \vspacerpostplot 
    \label{fig:ndust.ngas.m2}}
\end{figure}

\vspace{-0.5cm}
\section{Results}
\label{sec:results}

\subsection{Qualitative Behaviors: Critical Thresholds for Different Phenomena}
\label{sec:limits}

Figs.~\ref{fig:demo}-\ref{fig:demo.vs.alpha} show images of representative times during some of our simulations. It is clear that the dust and gas dynamics differ, sometimes dramatically.

Since here the dust does not act on gas, the gas dynamics are identical to those expected for supersonic MHD turbulence.  We confirm (for detailed analysis see \citealt{hopkins:gizmo}) that the highly super-sonic cases here ($\mathcal{M}\sim10$) develop a velocity scaling similar to the linewidth-size relation observed, with rms velocities on scale $\lambda$ of $\langle \mathcal{M}^{2}(\lambda) \rangle \sim \mathcal{M}(L_{\rm box})\,(\lambda/L_{\rm box})^{1/2}$, down to a sonic length $R_{\rm sonic}=\lambda(\mathcal{M}=1) \sim L_{\rm box}\,\mathcal{M}(L)^{-2}$, also as expected from previous work \citep{scalo:1998.turb.density.pdf,schmidt:2009.isothermal.turb,konstandin:2012.lagrangian.structfn.turb}. Below this scale, the turbulence becomes sub-sonic and we expect a \citet{kolmogorov:turbulence} cascade, $\mathcal{M}(\lambda<R_{\rm sonic}) \sim (\lambda/R_{\rm sonic})^{1/3}$. The gas forms a filamentary network of shocks and rarefactions; the characteristic width of filaments and dense structures is of order $R_{\rm sonic}$ \citep[see][]{vazquez-semadeni:1994.turb.density.pdf,padoan:1997.density.pdf,klessen:2000.pdf.supersonic.turb,kritsuk:2007.isothermal.turb.stats}. In our simulations with initially weak mean-field strengths, the initial magnetic fields grow exponentially until saturating with magnetic energy $\sim 5\%$ of the kinetic energy \citep[also as expected;][]{brandenburg:nonlinear.astro.dynamos,schekochihin:smallscale.turb.dynamo,federrath:supersonic.turb.dynamo}. As expected, for our driving routines, after a few dynamical times, the simulations reach a steady state, and the statistics of turbulent velocity fluctuations and dust dynamics do not evolve.

 \begin{figure}
    \plotonesize{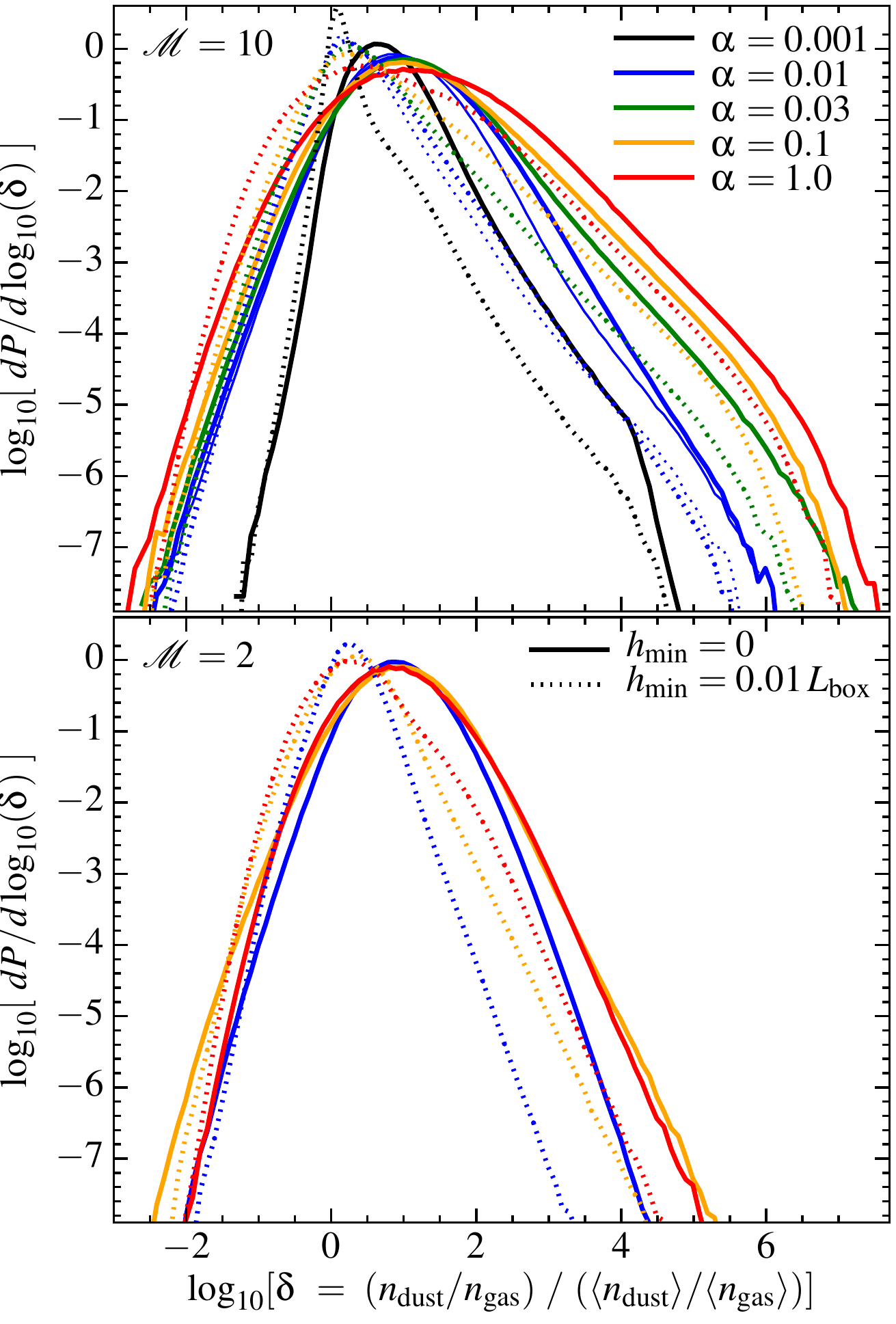}{0.9}
    \vspace{-0.25cm}
    \caption{Distribution of dust-to-gas ratios. We calculate $\delta$ (dust-to-gas ratio relative to mean) around each dust particle (at any gas density) in the distribution functions from Figs.~\ref{fig:ndust.ngas.m10}-\ref{fig:ndust.ngas.m2}, and plot the resulting PDF $dP/d\log(\delta)$ (time-averaged over the simulation). 
    {\em Top:} Mach number $\mathcal{M}\sim10$ cases from Fig.~\ref{fig:ndust.ngas.m10} (different $\sizeparam$ as labeled). As in the previous figure we also show the case with $\sizeparam=0.005-0.015$ as the thin line in the same style as the $\alpha=0.01$ case; the difference owing to a distribution of grain sizes of modest width is small. Solid lines show the case with $h_{\rm min}=0$; dotted lines show $h_{\rm min}=0.01\,L_{\rm box}$. For $h_{\rm min}>0$, the mean of the distributions is shifted by averaging closer to $\delta\sim1$, but the scatter is only reduced by a modest amount. The core of each distribution is approximately log-normal, but with a large ``tail'' towards high-$\delta$; this primarily arises in the low-density regions where the dust and gas de-couple (see Fig.~\ref{fig:dustgas.pvsweights}). 
    {\em Bottom:} Same, for the $\mathcal{M}=2$ cases. The scatter is reduced, primarily because the lower $\mathcal{M}$ means the dynamic range of {gas} densities is much smaller; this most noticeably suppresses the ``tail'' coming from very low $n_{\rm gas}$. 
    \vspacerpostplot 
    \label{fig:dustgas.pdf}}
\end{figure}

Note that the free-streaming length of a dust grain (relative to gas) can be estimated as $L_{\rm stream} \sim \langle |{\bf u}_{d}-{\bf u}_{\rm gas}| \rangle\,t_{s}$, where we expect (and confirm in our simulations) that the typical relative velocity $\langle |{\bf u}_{d}-{\bf u}_{\rm gas}| \rangle$ corresponds to the ``eddy velocity'' of turbulent modes on a scale $\sim L_{\rm stream}$ (much smaller modes do not strongly perturb the dust, and much larger coherent modes simply entrain both dust and gas together; see \citealt{voelk:1980.grain.relative.velocity.calc,ormel:2007.closed.form.grain.rel.velocities,pan:2010.grain.velocity.sims}). If we combine this with the expressions for $t_{s}$ and the velocity scalings above, we can approximate the solution as: 
\begin{align} 
\label{eqn:Lstream}
{L_{\rm stream}} &\sim 
\begin{cases}
	{\displaystyle \frac{\bar{\rho}_{d}\,a}{\rho_{\rm gas}} = \alpha\,\left(\frac{n_{\rm gas}}{\langle n_{\rm gas} \rangle}\right)^{-1}\,L_{\rm box}} \ \ \ \ \  \hfill { (L_{\rm stream} > R_{\rm sonic}) } \\
	{\displaystyle \, \, }\\
	{\displaystyle \, \, }\\
	{\displaystyle R_{\rm sonic}^{-1/2}\,\left(\frac{\bar{\rho}_{d}\,a}{\rho_{\rm gas}}\right)^{3/2} = \alpha^{3/2}\,\mathcal{M}\,\left(\frac{n_{\rm gas}}{\langle n_{\rm gas} \rangle}\right)^{-3/2}\,L_{\rm box}}\\
	{\displaystyle \, \, }\\
	 {\displaystyle \, \  \ } \hfill { (L_{\rm stream} < R_{\rm sonic}) }
\end{cases}
\end{align}
where the behavior differs depending on whether $L_{\rm stream}$ is above or below the sonic length (the motion is super or sub-sonic). Equating the two, we arrive at the critical $\alpha$ above which $L_{\rm stream} > R_{\rm sonic}$
\begin{align}
\alpha(L_{\rm stream} = R_{\rm sonic}) &= \mathcal{M}^{-2}\,\left(\frac{n_{\rm gas}}{\langle n_{\rm gas} \rangle}\right)^{-1}
\end{align}

\begin{figure}
    \plotonesize{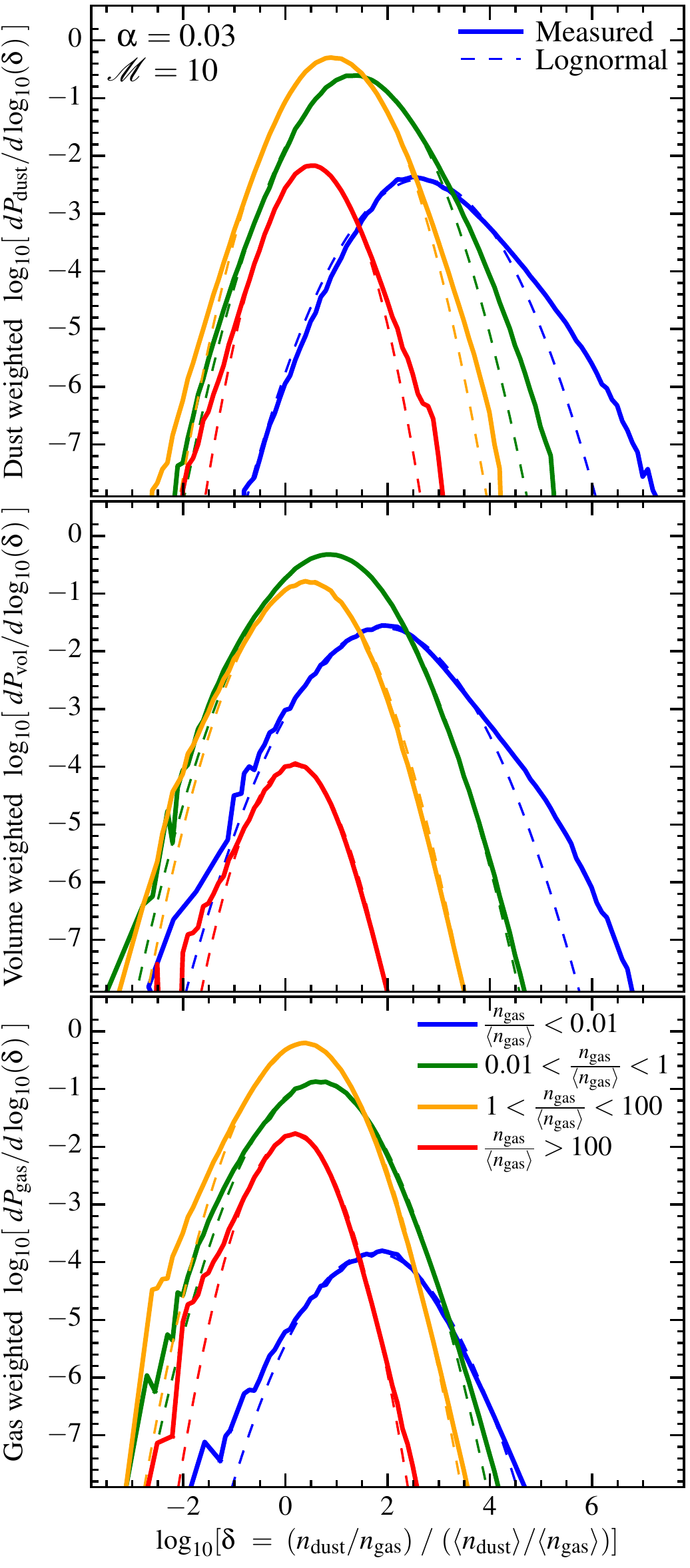}{0.9}
    \vspace{-0.25cm}
    \caption{Distribution of dust-to-gas ratios ($\delta$) at different gas densities $n_{\rm gas}$ (colors as labeled), for a single representative simulation ($\alpha=0.03$, $\mathcal{M}=10$, $\mathcal{M}_{A}=6$). {\em Top:} Distribution measured around each dust particle. At each $n_{\rm gas}$, the $\delta$ distribution is roughly log-normal (dashed line shows a log-normal with the same mean and variance), albeit with wider tails; this motivates our fitting function (Eq.~\ref{eqn:bivariate}). 
     The variance and mean depend on $n_{\rm gas}$; the high-$\delta$ tail in Fig.~\ref{fig:dustgas.pdf} comes from the weakly-coupled limit (low $n_{\rm gas}$), where $n_{\rm dust}\sim \langle n_{\rm dust} \rangle$ so $\langle \delta \rangle \gg1$. At high $n_{\rm gas}$, $\langle \delta \rangle \sim1$ but with large scatter.
     {\em Middle:} As {\em top}, except we sample $\delta$ around random points in space (the volume-weighted PDF, instead of the dust-mass weighted PDF; see \S~\ref{sec:fitfun}). The relative normalizations and median-$\delta$ values shift because low-density regions (with large volume-filling factors) are preferentially sampled, but the scatter is similar.
    {\em Bottom:} Same, except we sample $\delta$ around random gas elements, (the gas-mass weighted PDF). Median $\delta$ values shift closer to $\sim1$, but the scatter is similar.
    \vspacerpostplot 
    \label{fig:dustgas.pvsweights}}
\end{figure}

Consider the following limits: 

\begin{itemize}

\item{$\alpha\gtrsim1$: In isothermal strong shocks the maximum density enhancement is $\sim \mathcal{M}^{2}$; therefore, if $\alpha \gtrsim 1$, we expect $L_{\rm stream} \gtrsim L_{\rm sonic}$ even at these large post-shock densities, so dust can stream through even the densest structures assuming they have sizes of order the sonic length. The dust is therefore always near its mean density (little variance in $n_{\rm dust}$) while the gas dynamics proceed as usual (large variance in $n_{\rm gas}$). This is the weakly-coupled limit.}

\item{$1/\mathcal{M}^{2} \lesssim \alpha \lesssim 1$: The dust cannot ``break out'' of the most-dense structures. However, it can cluster on scales larger than the sonic length at the mean density. Therefore dust clustering is imprinted on the medium at intermediate densities and then turbulent compressions ``trap'' the fluctuations in dense regions. 

In very low-density regions where $n_{\rm gas} \lesssim \alpha\,\langle n_{\rm gas} \rangle$, $L_{\rm stream}\rightarrow L_{\rm box}$,  and the dynamics resemble the weakly-coupled case (although the low-density regions are precisely those where our neglect of Lorentz forces on dust is a poor approximation, and these may re-couple the dust and gas). 

In high density regions where $n_{\rm gas} \gtrsim \alpha\,\mathcal{M}^{2}\,\langle n_{\rm gas} \rangle$, the clustering scale drops below the sonic length, and the dynamics begin to resemble the sub-sonic case. In this limit, grains can be efficiently expelled from regions of high vorticity and trapped along lines of high strain, leading to their alignment in narrow filamentary structures \citep{cuzzi:2001.grain.concentration.chondrules,rouson:2001.grain.concentration.experiment,bec:2009.caustics.intermittency.key.to.largegrain.clustering,pan:2011.grain.clustering.midstokes.sims,monchaux:2012.grain.concentration.experiment.review}. Assuming the Kolmogorov scale is arbitrarily small, in this limit the maximum clustering amplitude becomes self-similar, because all grains ``see'' eddies which lie in an effectively infinite inertial range and are resonant with their own streaming timescale (so the clustering dynamics for grains of different sizes are simply rescaled in size; \citealt{hogan:2007.grain.clustering.cascade.model,yoshimoto:2007.grain.clustering.selfsimilar.inertial.range,bec:2008.markovian.grain.clustering.model}). Detailed discussion of this limit can be found in \citet{hopkins:2013.grain.clustering}; both experiments and numerical simulations of sub-sonic turbulence suggest that, provided infinite resolution, the small-scale dispersion in the dust-to-gas ratio converges to $\sim 0.4-0.5$\,dex.}

\item{$\alpha\ll 1/\mathcal{M}^{2}$: The dust is strongly coupled down to below the sonic scale, for all densities equal to or larger than the mean. This case is in the sub-sonic limit above over most of the domain. Since the dust is well-coupled at the mean density, gas being compressed into structures of order the sonic scale has approximately the mean dust-to-gas ratio, with little scatter. So although the sub-sonic processes described above can operate, the variance we expect at intermediate to high densities is smaller because there are no ``seed'' fluctuations in the dust-to-gas ratio in the diffuse medium.}

\end{itemize}

Each of these behaviors are illustrated in Figs.~\ref{fig:demo}-\ref{fig:demo.vs.alpha}. Table~\ref{tbl:sims} gives a complete list of the simulations, and summarizes some of their salient properties, which we will discuss below.

\begin{figure}
    \plotonesize{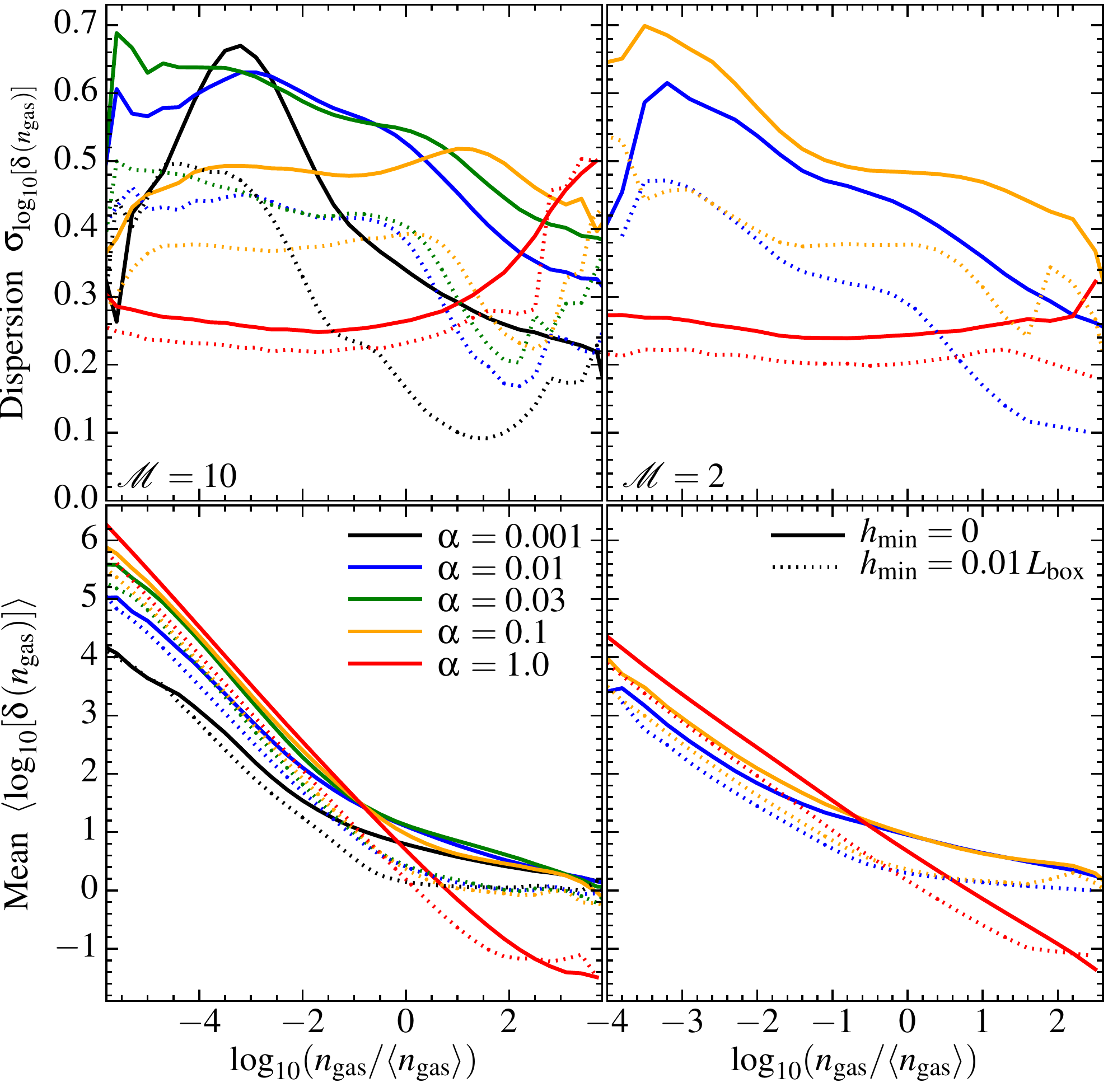}{1.02}
    \vspace{-0.25cm}
    \caption{{\em Top:} $1\sigma$ logarithmic dispersion in the dust-to-gas ratio $\log_{10}(\delta)$ (or equivalently dust density $\log_{10}(n_{\rm dust})$) at fixed gas density $n_{\rm gas}$ (see Eq.~\ref{eqn:bivariate}). We show the simulations from Fig.~\ref{fig:ndust.ngas.m10} with $\mathcal{M}\sim10$ ({\em left}) and those from Fig.~\ref{fig:ndust.ngas.m2} with $\mathcal{M}\sim2$ ({\em right}). For both, we show the results smoothed within a radius $h_{\rm min}=0$ ({\em solid}) and $h_{\rm min}=0.01\,L_{\rm box}$ ({\em dotted}). The variance is generally maximized for $\alpha\sim0.01-0.1$, but is generally only weakly dependent on $\alpha$ and $n_{\rm gas}$. Smoothing on larger scales ($h_{\rm min}\sim 0.01\,L_{\rm box}\sim R_{\rm sonic}$) decreases the dispersion, but only by a modest $\sim 0.1$\,dex. 
    {\em Bottom:} Logarithmic mean $\langle \log_{10}(\delta) \rangle$ in the dust-to-gas ratio relative to mean ($\delta$) as a function of gas density. For $\alpha\lesssim1$, at low densities ($n_{\rm gas}\lesssim \alpha \langle n_{\rm gas} \rangle$), the dust de-couples from the gas so the mean $\langle \delta \rangle \propto n_{\rm gas}^{-1}$; at high densities the dust and gas couple more tightly so $\langle \delta \rangle \rightarrow 1$. For $\alpha \gtrsim 1$ the dust streams through sonic-length structures and remains relatively poorly coupled until much higher densities (see \S~\ref{sec:limits}).
    \vspacerpostplot 
    \label{fig:variance}}
\end{figure}

\vspace{-0.5cm} 
\subsection{The Distribution of Dust and Gas Densities}


We now consider these effects more quantitatively. In Figs.~\ref{fig:ndust.ngas.m10}-\ref{fig:ndust.ngas.m2}, we measure the density of both dust and gas around every dust particle in the simulation, averaged on a smoothing scale $h_{\rm min}$,\footnote{We use a standard kernel-density estimator to calculate the density of dust particles and gas particles around each point ${\bf x}_{i}$ based on the distribution of dust/gas particle neighbors, e.g.\ $n_{{\rm dust},\,i} = \sum_{j} W({\bf x}_{j}-{\bf x}_{i},\,h_{i})$, where $h = {\rm MAX}(h_{\rm min},\,h_{N})$ with $h_{N}$ the radius that encloses a finite neighbor number of particles $j$, and $W$ the kernel function chosen here for consistency to match the same used in our mesh-free hydrodynamics methods. This is much more accurate, given the Lagrangian nature of our code, than a simpler particle-in-cell estimate. We have confirmed that our results are insensitive to the (arbitrary) neighbor number in the estimation kernel and the specific choice of kernel function. We have also confirmed that a direct reconstruction output by our hydrodynamic solver gives indistinguishable results to this estimator applied in post-processing.} and plot the normalized 2D histogram of all points in the $n_{\rm gas}-n_{\rm dust}$ plane. Since, as noted above, the turbulence becomes steady-state after $\sim1$ crossing time, the results at any individual time output are statistically identical; we therefore simply combine all outputs after the first few crossing times to reduce the sampling noise. 

As expected from our arguments above, at sufficiently low $n_{\rm gas}$ and large $\alpha$, the dust de-couples from the gas, residing at the mean dust density independent of the gas density, with small fluctuations. At higher densities, the dust tracks the gas on average, $\langle n_{\rm dust}(n_{\rm gas}) \rangle \propto n_{\rm gas}$, but with obvious scatter. Note the scatter is much larger than the Poisson noise and is numerically converged (see Appendix~\ref{sec:resolution}).



Because the dust does not alter the gas dynamics in our simulations, the bivariate distribution $P(n_{\rm gas},\,n_{\rm dust})$ is separable into $P(n_{\rm gas})\,P(n_{\rm dust}\,|\,n_{\rm gas})$. The distribution of gas density $P(n_{\rm gas})$ in isothermal MHD turbulence is well-studied, and we confirm the usual log-normal form \citep{passot:1988.proof.lognormal,vazquez-semadeni:1994.turb.density.pdf,molina:2012.mhd.mach.dispersion.relation,konstantin:mach.compressive.relation,federrath.2015:density.pdf.and.sfr.in.polytropic.turbulence} with subtle non-lognormal deviations consistent with those predicted in \citet{hopkins:2012.intermittent.turb.density.pdfs} and confirmed in \citet{federrath:2013.intermittency.vs.numerics}. The interesting behavior we wish to study here is encapsulated in the non-universal dust-to-gas ratio $P(n_{\rm dust}\,|\,n_{\rm gas})$.


Figs.~\ref{fig:dustgas.pdf} therefore collapses our 2D distribution functions into the distribution of dust-to-gas ratio, which we define for convenience relative to the mean in the box:
\begin{align}
\delta \equiv \frac{ n_{\rm dust}/ n_{\rm gas}} {\langle n_{\rm dust} \rangle / \langle n_{\rm gas} \rangle}
\end{align}
We find in every case a broad distribution, with a log-normal ``core'' and low-$\delta$ behavior, and a power-law like tail at the highest $\delta$. 

The origin of this behavior is demonstrated in Fig.~\ref{fig:dustgas.pvsweights}, which plots $\delta$ at fixed $n_{\rm gas}$; this is equivalent to $P(n_{\rm dust}\,|\,n_{\rm gas})$. At a given gas density, $\delta$ (or $n_{\rm dust}$) is distributed approximately log-normally. The dispersion $\sigma$ of the log-normal depends relatively weakly on $n_{\rm gas}$, while the mean $\langle\ln(\delta)\rangle_{\rm dust}$ shifts systematically. At low gas densities ($n_{\rm gas} \lesssim \alpha\,\langle n_{\rm gas} \rangle$) there is some excess at high-$\delta$ with respect to a log-normal fit, but most of the deviation from a log-normal in Fig.~\ref{fig:dustgas.pdf} arises because it represents an integral over the different log-normal $P(n_{\rm dust}\,|\,n_{\rm gas})$ with different mean values.

A physical argument for why the dust-to-gas ratio should be distributed log-normally in the high-density limit is presented in \citet{hopkins:2013.grain.clustering}. Essentially, each encounter between a Lagrangian ``parcel'' of grains and a vorticity/strain structure or eddy in the turbulence imparts an essentially random multiplicative factor on the local grain density (the factor depends on the magnitude of the vorticity/strain, and orientation of the eddy relative to the grain velocity). Integrating over time and structures of a wide range of sizes, this random multiplicative process produces a quasi-lognormal distribution. Unlike the gas density fluctuations (where log-normal behavior is specific to isothermal gas) this expectation is independent of the gas equation of state, and has been seen in sub-sonic experiments with strictly adiabatic, incompressible gas \citep{hogan:1999.turb.concentration.sims,bec:2007.grain.clustering.markovian.flow,pan:2011.grain.clustering.midstokes.sims}. 

Predicting the magnitude of the fluctuations $\sigma$ is more challenging. Again, some analytic arguments are presented in \citet{hopkins:2013.grain.clustering}: they show that a single encounter with a ``resonant'' structure with eddy crossing time $\sim t_{\rm stop}$ and coherence length $\sim L_{\rm stream}$ leads to an approximate factor $\sim 2$ ($0.3\,$dex) multiplicative effect on the dust-to-gas ratio. Larger/smaller structures produce weaker effects. Broadly similar multiplicative effects occur when dust grains pass through a shock or rarefaction with gas velocity gradient $|\nabla\cdot {\bf v}| \sim 1/t_{\rm stop}$ \citep[see e.g.][]{booth:dust.gas.2d.shock.tests,loren:two.fluid.dusty.gas.sims}. Assuming the dispersion is dominated by a couple such encounters per global grain-crossing time, values $\sigma \sim 0.3-0.6$\,dex (seen in Fig.~\ref{fig:dustgas.pvsweights}) are plausible. However, the magnitude of this effect depends on the geometry and filling factor of structures in the turbulence, so more detailed analytic models are needed.

\vspace{-0.5cm}
\subsubsection{Approximate Fitting Functions}
\label{sec:fitfun}

Based on the above, the bivariate distribution of dust and gas densities (around a random dust particle) in Figs.~\ref{fig:ndust.ngas.m10}-\ref{fig:ndust.ngas.m2} can be approximated by: 
\begin{align}
\nonumber \frac{dP_{\rm dust}}{d\ln{n_{\rm gas}}\,d\ln{n_{\rm dust}}} &=
\frac{1}{2\pi\sqrt{S_{\rm gas}\,S_{\rm dust}}}\,
\exp{\left[ -\left(\frac{\Delta_{\rm gas}^{2}}{2\,S_{\rm gas}} + \frac{\Delta_{\rm dust}^{2}}{2\,S_{\rm dust}}\right) \right]}\\
\nonumber S_{\rm dust} &= S_{\rm dust}(n_{\rm gas},\,h_{\rm min}) = \left( \sigma_{\log_{10}[\delta(n_{\rm gas})]}\,\ln{10} \right)^{2}\\
\nonumber \Delta_{\rm dust} &\equiv \ln{(\delta)} - \langle \ln{\delta(n_{\rm gas})} \rangle_{\rm dust}\\
\label{eqn:bivariate} \Delta_{\rm gas} &\equiv \ln{\left(\frac{n_{\rm gas}}{\langle n_{\rm gas}\rangle}\right)} + \frac{S_{\rm gas}}{2}
\end{align}
where $S_{\rm gas} = S_{\rm gas}(\mathcal{M})$ is a constant for a given simulation (given by the usual relations in super-sonic isothermal turbulence; see Appendix~\ref{sec:alt.pdfs}), while $S_{\rm dust}$ and $\langle \ln{\delta(n_{\rm gas})} \rangle_{\rm dust}$ are the variance and mean in $\ln{\delta}$ {\em at a given $n_{\rm gas}$}. 

Fig.~\ref{fig:variance} plots the variables in Eq.~\ref{eqn:bivariate}, as a function of $n_{\rm gas}$. Specifically we measure the logarithmic mean $\langle \log_{10}(\delta) \rangle = \langle \ln{(\delta)} \rangle/\ln{10}$ and dispersion $\sigma_{\log_{10}[\delta(n_{\rm gas})]}$ in $\delta(n_{\rm gas})$. 

The dispersion/variance in $\log{\delta}$ is maximized (at most $n_{\rm gas}$) for $\alpha\sim 0.01-0.1$. It drops for $\alpha \lesssim 0.001$ (when the grains are strongly-coupled so track gas closely), and $\alpha\gtrsim1$ (when the grains are weakly-coupled, so stay near $\langle n_{\rm dust} \rangle$). In general the dispersion decreases weakly with higher $n_{\rm gas}$ as grains become more tightly-coupled (except $\alpha\gtrsim1$, where the grains are very weakly coupled so only begin to cluster at the highest densities). But in all cases the dependence of $\sigma_{\log_{10}[\delta(n_{\rm gas})]}$ on $n_{\rm gas}$ is weak, with typical values $\sim 0.3-0.6\,$dex ($S_{\rm dust}\sim 0.5-2$). The mean $\langle \log_{10}(\delta) \rangle$ shows a clear transition: at low $n_{\rm gas}$, in the weakly-coupled regime, dust resides near $\langle n_{\rm dust} \rangle$ so $\langle \delta \rangle \sim n_{\rm gas}^{-1}$, while at high $n_{\rm gas}$, in the tightly-coupled regime, $\langle \delta \rangle \rightarrow 1$.

Because our default calculation measures the properties around each dust particle, this is the dust-mass-weighted PDF $dP_{\rm dust}$. We could, instead, uniformly sample the volume (giving the volume-weighted PDF $dP_{\rm vol}$) or the gas mass $dP_{\rm gas}$. The differences between each are discussed in detail in Appendix~\ref{sec:alt.pdfs}; for the pure point-wise PDF ($h_{\rm min}\rightarrow 0$), they are all trivially related: $dP_{\rm gas} \propto n_{\rm gas}\,dP_{\rm vol} \propto (n_{\rm gas}/n_{\rm dust})\,dP_{\rm dust}$. As expected, then, volume-weighted PDF shifts the weight towards low-density regions, which occupy a larger volume, while the gas-mass weighted PDF shifts the weight towards higher gas densities and gas-to-dust ratios, bringing the average $\delta$ closer to unity. However, it is easy to show for Eq.~\ref{eqn:bivariate} that these transformations preserve the lognormal shape of the PDF, and do not change the variance $S_{\rm dust}$ or $S_{\rm gas}$. They simply shift the mean values of $\Delta_{\rm dust}$ and $\Delta_{\rm gas}$.\footnote{For $dP_{\rm vol}$, we have Eq.~\ref{eqn:bivariate} with $\langle \ln{(\delta)} \rangle_{\rm vol} = \langle \ln{\delta(n_{\rm gas})} \rangle_{\rm dust} - S_{\rm dust}$. For $dP_{\rm gas}$, we apply the same shift in $\langle \ln{(\delta)} \rangle$ and also shift the mean $\langle \ln(n_{\rm gas}/\langle n_{\rm gas} \rangle ) \rangle_{\rm gas} = +S_{\rm gas}/2$. See Appendix~\ref{sec:alt.pdfs} for details.}

\begin{figure}
    \vspace{-0.1cm}
  \includegraphics[width=0.99\columnwidth]{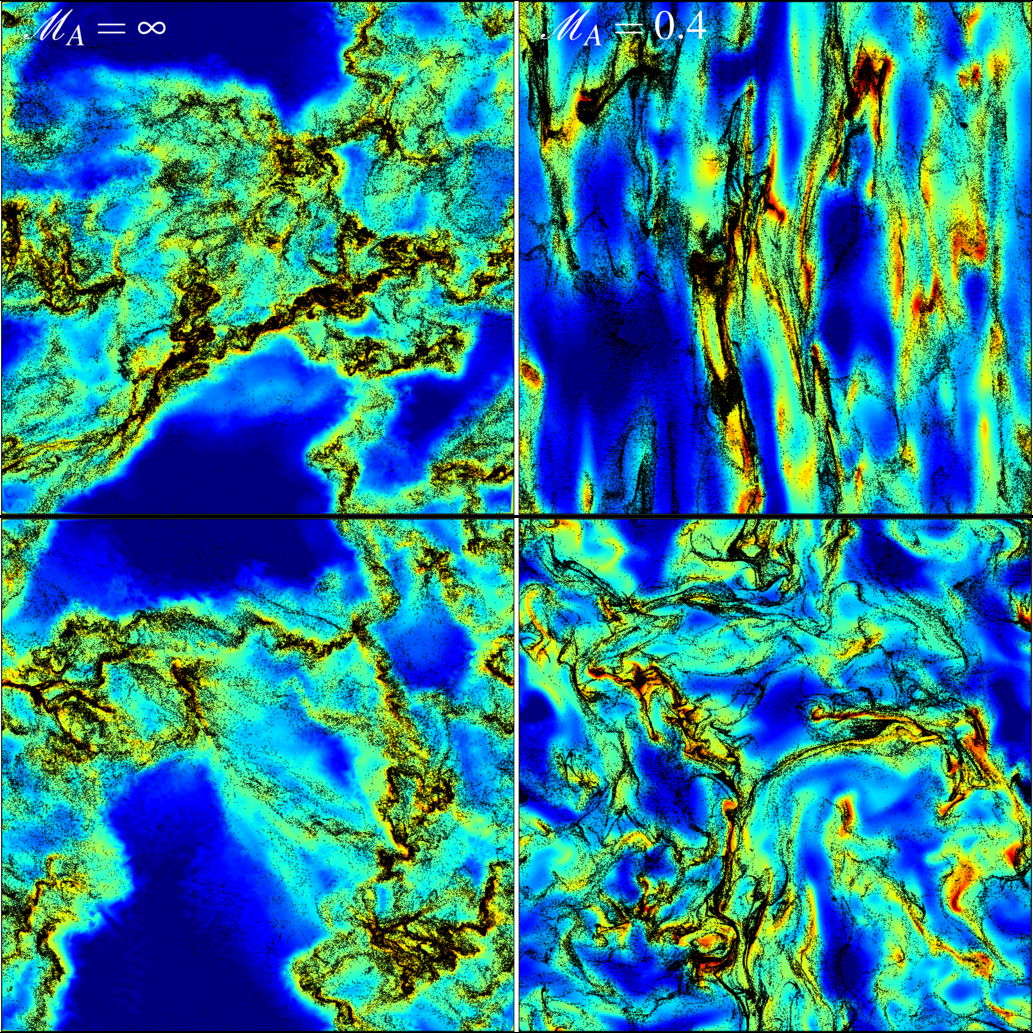} 
    \vspace{-0.25cm}
    \caption{Image of dust and gas (as Fig.~\ref{fig:demo.vs.alpha}) in simulations with $\mathcal{M}\sim10$ and $\alpha=0.03$ but varied magnetic mean-field strength: no magnetic field ($\mathcal{M}_{A}=\infty$; {\em left}) or very strong mean-fields ($\mathcal{M}_{A}=0.4$; {\em right}). {\em Top:} $x-z$ projection, where $\hat{z}$ is the mean-field direction. {\em Bottom:} $x-y$ projection. The strong-field case clearly produces global anisotropy and more filamentary structure; it also slightly enhances the segregation of dust and gas.
    \vspacerpostplot 
    \label{fig:bfield.image}}
\end{figure}

\vspace{-0.5cm}
\subsubsection{Dependence on Smoothing Scale and Mach Number}

Note in Figs.~\ref{fig:ndust.ngas.m10}-\ref{fig:ndust.ngas.m2}, we consider three values of $h_{\rm min}$. First, $h_{\rm min}=0$, i.e.\ the fluctuations measured on the smallest possible (resolution) scale. Then, $h_{\rm min}=0.001\,L_{\rm box}$ and $h_{\rm min}=0.01\,L_{\rm box}$. These give the {\em average} density calculated around each point, averaged within a finite volume-averaging spherical radius equal to $h_{\rm min}$. Recall, for our $\mathcal{M}=10$ simulations, $h_{\rm min}=0.01\,L_{\rm box}$ corresponds to the sonic length $R_{\rm sonic}$. As must occur, the variation in $n_{\rm dust}/n_{\rm gas}$ decreases as $h_{\rm min}$ increases (obviously, in the limit $h_{\rm min}\rightarrow L_{\rm box}$, all points collapse to exactly the box-averaged $\langle n_{\rm dust} \rangle$ and $\langle n_{\rm gas} \rangle$). For smaller values of $\alpha$, the variance in $n_{\rm dust}$ at fixed $n_{\rm gas}$ decreases more rapidly as we increase $h_{\rm min}$ -- this is because, as noted above, the physical scale of the clustering is smaller for smaller $\alpha$ (all else being equal), so by smoothing on a fixed scale $h_{\rm min}$ we are averaging-out more of the small-scale variations. Still, for most of our simulations, significant variation persists even with $h_{\rm min}=0.01$. 

At low Mach numbers (our $\mathcal{M}=2$ suite), we robustly find that the variance in $\log{\delta}$ at high $n_{\rm gas}/\langle n_{\rm gas} \rangle$ is lower for a given $\alpha$. This is not surprising. For one, the typical magnitude of gas density fluctuations is smaller in these cases, which in turn generates smaller ``seed'' fluctuations for dust. More importantly, recall, in the sub-sonic limit, $L_{\rm stream} \propto \mathcal{M}$ at fixed $\alpha$ and $n_{\rm gas}/\langle n_{\rm gas} \rangle$, while $\alpha_{s} \propto \mathcal{M}^{-2}$ -- in other words, the free-streaming length of the dust relative to the box size, and relative to the sonic length ($L_{\rm stream}/L \propto \mathcal{M}^{3}$), are lower by factors of $\sim 5$ and $\sim 125$, respectively, in our $\mathcal{M}\sim 2$ cases. Another way to think of this is to simply note that, in {\em physical} units (at fixed $\alpha$ and assuming clouds lie on the local linewidth-size relation), our $\mathcal{M}\sim2$ runs correspond to $L_{\rm box}\sim0.4\,$pc instead of $\sim 10\,$pc at $\mathcal{M}=10$, and correspondingly the physical grain size $a_{d}$ is a factor $\sim 25$ smaller. So it is not, in fact, physical to think of our $\mathcal{M}=2$ runs as ``the same'' physical setup as $\mathcal{M}=10$ with simply a different Mach number -- they are measuring grain fluctuations corresponding to different physical grain sizes on much smaller physical scales. Given this, the remarkable fact is how {\em similar} the trends are, implying that the dependence of the dust-to-gas fluctuations on scale, turbulent properties, and grain size are surprisingly weak.

\begin{figure}
 \begin{tabular}{c}
  \vspace{0.1cm}
  \includegraphics[width=0.97\columnwidth]{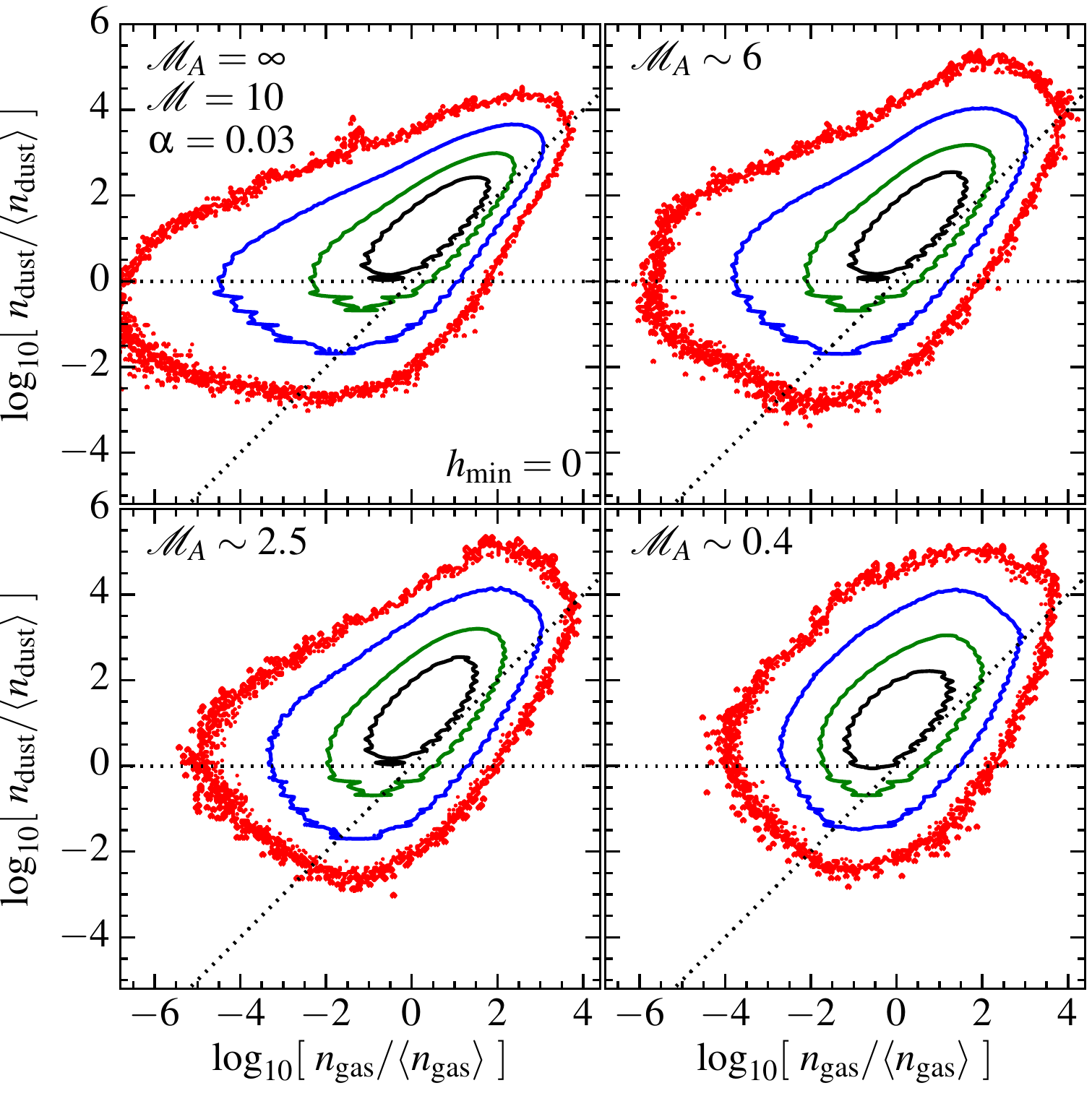} \\
  \includegraphics[width=0.97\columnwidth]{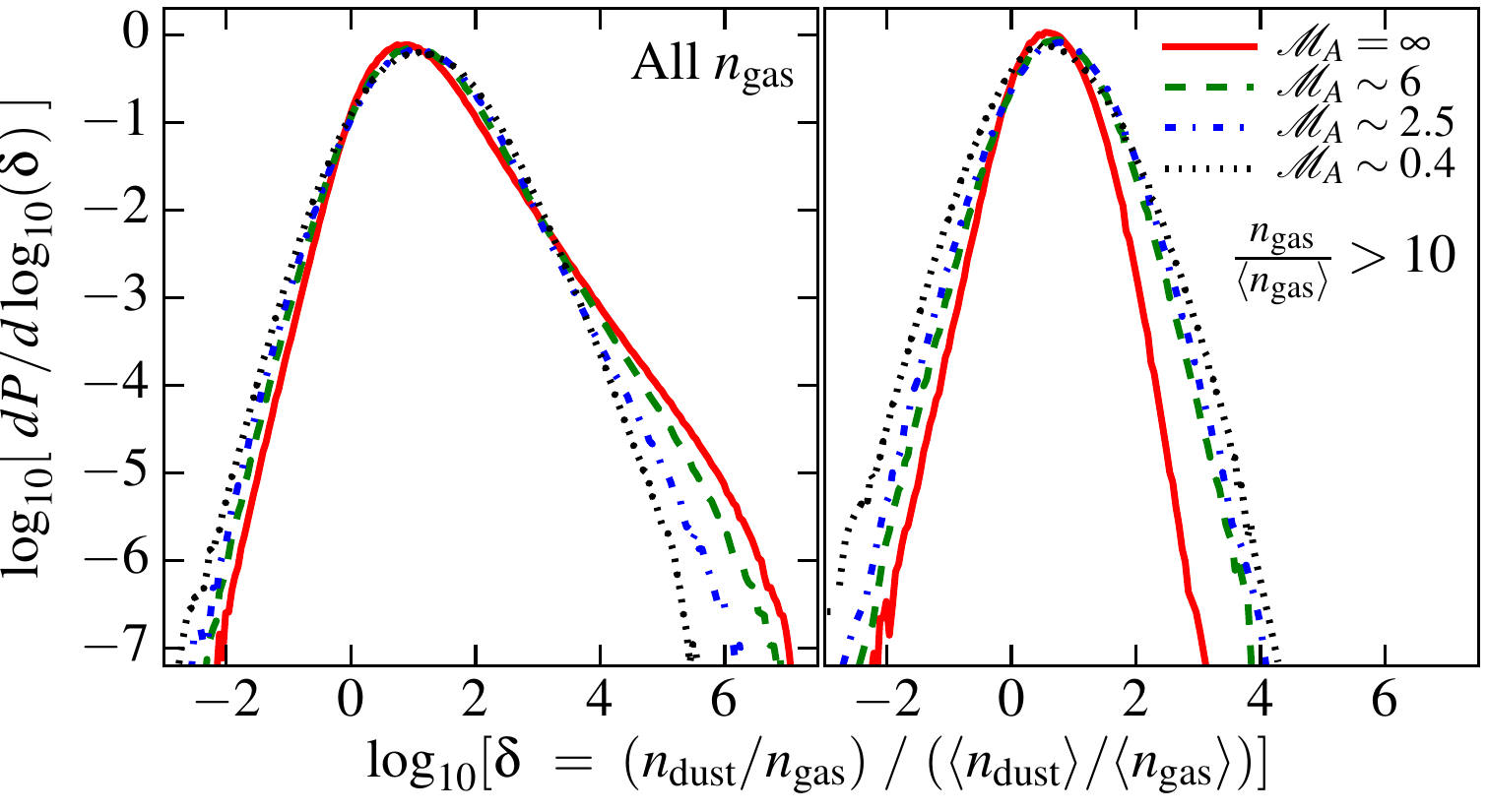} 
 \end{tabular}
    \vspace{-0.25cm}
    \caption{{\em Top:} Bivariate distribution of dust and gas densities as Fig.~\ref{fig:ndust.ngas.m10}, for the $\mathcal{M}\sim10$, $\alpha=0.03$ simulations where we vary the magnetic mean-field strength (see Table~\ref{tbl:sims}). {\em Bottom:} Distribution of dust-to-gas ratios $\delta$ as Fig.~\ref{fig:dustgas.pvsweights}, for all gas ({\em left}) and just the high-density gas ({\em right}). Increased magnetic field strengths suppress the low-density tail of {\em gas} density fluctuations, where the dust and gas de-couple, so the highest $\delta$ values arising from these low-density regions decrease. However, at a fixed density $n_{\rm gas}$, increasing field strength (lower $\mathcal{M}_{A}$) actually {\em increases} the dust-to-gas ratio fluctuations. The effect is weak compared to variations in $\alpha$, but true at every $n_{\rm gas}$ we measure here. This owes to the increased solenoidal/vorticity component of the flows with stronger fields (which produces dust density fluctuations without corresponding gas density fluctuations) and to the existence of additional magnetic pressure to create local ``pressure traps.'' 
    \vspacerpostplot 
    \label{fig:bfield.effects}}
\end{figure}

\vspace{-0.5cm}
\subsubsection{Dependence on Mean-Field Strength}

Thus far, we have focused on simulations with small {mean} (coherent box-scale) magnetic fields. This simplifies our study as the saturated field depends only on the sonic Mach number $\mathcal{M}$. However, this does not mean the fields are negligible: in our $\mathcal{M}\sim10$ simulations, the rms field strength in physical units is $\langle |{\bf B}|^{2} \rangle^{1/2} \sim 4.2\,\mu G \,\sqrt{(\langle n_{\rm gas} \rangle /10\,{\rm cm^{-3}})\,(T/100\,{\rm K})}$, comparable to observations \citep[see e.g.][]{elmegreen:2004.obs.ism.turb.review,brown:galactic.b.field,crutcher:cloud.b.fields} in typical clouds. 

Even absent Lorentz forces on dust, magnetic fields can alter concentration. For example, fields modify the velocity scalings,  imprint local anisotropy, and change the ratio of solenoidal to compressible modes \citep[see e.g.][]{kowal:2007.log.density.turb.spectra,burkhart:2009.mhd.turb.density.stats,lemaster:2009.density.pdf.turb.review,kritsuk:2011.mhd.turb.comparison,molina:2012.mhd.mach.dispersion.relation,downes:2012.multifluid.turb.density.pdf,federrath:2012.sfe.pwrspec.vs.time.sims}. All of these can change the ``response'' of a parcel of grains to an encounter with a turbulent velocity structure \citep[][]{lazarian:2004.mhd.effects.dont.dramatically.change.turb.concentration,yan.2004:lorentz.forces.drag.dust.ism.analytic,hopkins:2013.turb.planet.direct.collapse,hopkins:2013.grain.clustering,hopkins:2014.pebble.pile.formation}. But these effects depend not just on the rms field strength but also on the mean-field strength, especially as the turbulence goes from super-Alfv{\'e}nic to sub-Alfv{\'e}nic \citep[][and references therein]{collins.2012:trans.alfvenic.mhd.turb.cloud.transition.states}. 

We therefore consider simulations with $\alpha=0.03$ and $\mathcal{M}\sim 10$ fixed\footnote{Non-linear interactions make it difficult to maintain exactly $\mathcal{M}=10$ as we vary $|\langle {\bf B} \rangle |$; however the runs in Table~\ref{tbl:sims} all saturate with $\mathcal{M}\approx 8-12$.} and varying mean-field $|\langle {\bf B} \rangle |$. We consider our ``default'' case above ($|\langle {\bf B} \rangle |\approx 0.5$, with mean Alfv{\'e}n $\mathcal{M}_{A}\sim 6$), a pure hydro case ($|\langle {\bf B} \rangle |=0$, $\mathcal{M}_{A}=\infty$), and two strong-field cases with $|\langle {\bf B} \rangle | = (4.3,\,27)$, producing saturated $\mathcal{M}_{A} \approx (2.5,\,0.4)$, i.e.\ rms field strengths $\langle |{\bf B}|^{2} \rangle^{1/2} \sim (10,\,65)\,\mu G \,\sqrt{(\langle n_{\rm gas} \rangle /10\,{\rm cm^{-3}})\,(T/100\,{\rm K})}$.

Figs.~\ref{fig:bfield.image}-\ref{fig:bfield.effects} show the results. Most obviously, increasing the field strength suppresses gas-density fluctuations, especially at very low densities $n_{\rm gas} \ll \langle n_{\rm gas} \rangle$ (where magnetic fields dominate the pressure); this effect is well-known \citep[see][]{ostriker:2001.gmc.column.dist,lemaster:2009.density.pdf.turb.review,burkhart:2009.mhd.turb.density.stats,collins.2012:trans.alfvenic.mhd.turb.cloud.transition.states}. Because the dust and gas de-coupled at low densities, this can weakly reduce the absolute magnitude of dust concentration (e.g.\ the clumping factor) averaged over all densities in the simulation.

However, {\em at fixed density} ($n_{\rm gas}$), variations in the dust-to-gas ratio $\delta$ are {\em stronger} with larger field strengths. Stronger fields direct more energy into solenoidal as opposed to compressive modes (evident in the coherent filamentary structure and smaller voids in Fig.~\ref{fig:bfield.effects} for $\mathcal{M}_{A}=0.4$); the solenoidal (vorticity/strain) modes can still induce large changes in dust density (see references in \S~\ref{sec:intro}), but do not alter the gas density, so they directly alter the dust-to-gas ratio. Moreover, magnetic fields provide another source of pressure which the grains do not feel -- this can produce phenomena such as zonal flows which create ``pressure traps'' (local maxima) in which grains with  appropriate stopping times collect \citep[see][]{whipple:solid.concentration.pressure.traps,pinilla:dust.trapping.zonal.flow.instabilities.sam,dittrich:2013.grain.clustering.mri.disk.sims}. The effects are weak compared to changing $\alpha$, but are significant at every density $n_{\rm gas}$.

We caution, however, that we have neglected Lorentz forces on grains, which of course become larger with increasing field strength, and may therefore reverse some of these effects.

\begin{figure}
    \plotonesize{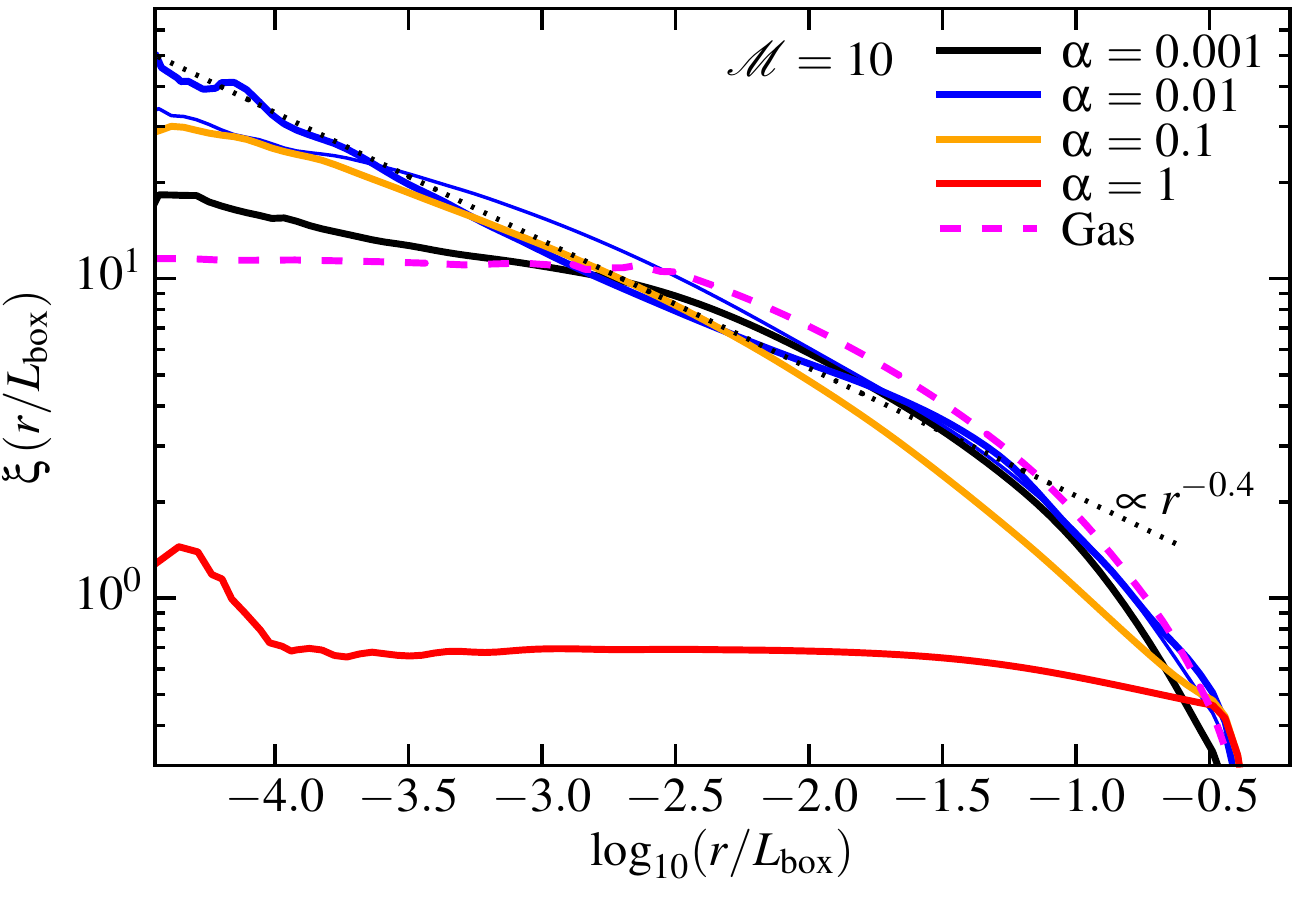}{0.97}
    \vspace{-0.25cm}
    \caption{The auto-correlation functions of the dust ($\xi_{\rm dust}$) and gas ($\xi_{\rm gas}$). Here we measure the three-dimensional, isotropic, radial $\xi(r)$, per Eq.~\ref{eqn:corrfn} (at single random instant in time); this relates to the radial distribution function $g(r)=1+\xi(r)$. This measures the variance in density fluctuations ($\langle \delta(r)\rangle^{2}/\langle\delta\rangle^{2}$) smoothed on different scales. The gas $\xi_{\rm gas}$ flattens below a scale of order the sonic length $\sim R_{\rm sonic} \sim L_{\rm box}/\mathcal{M}^{2}$, as expected. The dust clustering is reflected by $\xi_{\rm dust}$ continuing to rise to much smaller scales. In all cases with $\alpha\ll1$, $\xi_{\rm dust} \propto r^{-0.3-0.5}$ over the clustering dynamic range (down to the free-streaming scale $L_{\rm stream}$ at the maximum gas densities). For $\alpha\gtrsim1$ there is little clustering, except in the most extreme high-density regions (the ``bump'' at small scales). 
    \vspacerpostplot 
    \label{fig:corrfn}}
\end{figure}

\vspace{-0.5cm}
\subsubsection{Dust and Gas Correlation Functions}

The scale-dependence of grain clustering is also reflected in the autocorrelation function $\xi(r)$, defined in the usual fashion as
\begin{align}
\label{eqn:corrfn}
\xi(r)_{\rm dust} &\equiv \frac{1}{\langle n_{\rm dust} \rangle}\,\left\langle \frac{d N_{\rm dust}({\bf r})}{d^{3}{\bf x}} \right\rangle_{|{\bf r}|=r} - 1
\end{align}
where $\langle {d N_{\rm dust}({\bf r})}/{d^{3}{\bf x}} \rangle$ is the average number density of dust particles at a distance $r=|{\bf r}|$ from each dust particle.\footnote{Note in the terrestrial literature, it is more common to define the radial distribution function $g(r)$, but this is trivially related to $\xi(r)$ by $g(r) = 1 + \xi(r)$.} Replacing dust with gas, we have the autocorrelation of gas (mass), $\xi(r)_{\rm gas}$. These are particularly simple to compute given our Lagrangian numerical method. In cases with a weak mean (coherent) magnetic field, there is no preferred direction in our simulations, so we need consider only the isotropic $\xi(r)$ (even in the strong-mean field case we simulate, the anisotropic corrections are small compared to the effects of different grain sizes). Fig.~\ref{fig:corrfn} shows the correlation function for dust and gas in our $\mathcal{M}\sim10$, $\mathcal{M}_{A}\sim6$ simulations.

The gas correlation function for all our runs with the same $\mathcal{M}$ (and $\mathcal{M}_{A}$) is identical (since the gas dynamics are not altered by the dust) and behaves as expected. $\xi_{\rm gas}$ rises towards small scales, until it flattens completely at a scale $r$ a factor of a few below the sonic scale $R_{\rm sonic}\sim L_{\rm box}/\mathcal{M}^{2}$, where pressure effects suppress small-scale density fluctuations.

At large scales $\gtrsim({\rm a\ couple})\, \alpha\,L_{\rm box} \sim \langle L_{\rm stream} \rangle$, the dust is well-coupled to the gas (at least around the mean $n_{\rm gas}$), so $\xi_{\rm dust}\sim \xi_{\rm gas}$. But below this scale the two de-couple. Initially (scales $R_{\rm sonic} \lesssim r \lesssim \langle L_{\rm stream} \rangle$), the dust clusters more weakly than the gas ($\xi_{\rm dust} < \xi_{\rm gas}$). This reflects the dust free-streaming. But dust clustering/trapping in dense gas means $\xi_{\rm dust}$ continues to rise as a power-law towards much smaller scales $\sim L_{\rm stream}(n_{\rm gas}^{\rm max})$ (where $n_{\rm gas}^{\rm max}$ represents the highest gas densities reached by a significant volume fraction, $\sim 100\,\langle n_{\rm gas} \rangle$ in these runs). We can approximate $\xi_{\rm dust}$ reasonably well over most of its dynamic range with a power-law  
\begin{align}
\xi_{\rm dust} = \left(\frac{r}{r_{0}} \right)^{-\eta}
\end{align}
with $\eta\approx 0.3-0.5$ (depending on the simulation) and $r_{0}\sim (0.3-1)\, L_{\rm box}$ (note $\xi$ must drop more rapidly and eventually become negative as $r\rightarrow L_{\rm box}$, but this is not interesting). For a power-law $\xi(r)$, the variance in the density field averaged within a spherical radius $r$ (or equivalently, the mass enclosed in spheres of radius $r$), is trivially related to $\xi$ by
\begin{align}
\sigma^{2}_{n_{\rm dust}}(r) \equiv \frac{\langle n_{\rm dust}(r)^{2} \rangle}{\langle n_{\rm dust} \rangle^{2}} = C_{\eta}\,\left( \frac{r}{r_{0}} \right)^{-\eta}
\end{align}
where $C_{\eta}\approx 1.035\sim1$ for all $\eta$ in the range of interest \citep{peebles:1993.cosmology.textbook}. So the rms dispersion in $n_{\rm dust}$ scales as $\sigma_{n_{\rm dust}} \approx (r/L_{\rm box})^{-\eta/2} \sim  (r/L_{\rm box})^{-0.2}$. The small value of $\eta/2\approx 0.2$ here means that the scale-dependence of the fluctuations is weak; this is why we saw relatively mild changes in the PDF as we increased $h_{\rm min}$.\footnote{Note that $\xi(r)$ is three-dimensional. The projected/angular correlation function $\omega(\theta)$ is simply related by integration. Because $\xi(r) \propto r^{-\eta}$ is sufficiently shallow ($\eta<1$), we obtain a nearly-flat projected $\omega(\theta)\sim\,$constant.}

Note, however, that the scaling $\xi(r)$ determines how the {\em volume-weighted, linear} density variance scales (not logarithmic variance); moreover it measures this for $n_{\rm dust}$ not the dust-to-gas ratio (so some of the power comes from gas-density fluctuations). Nonetheless since the power in gas-density fluctuations is small on small scales, this should translate there to dust-to-gas fluctuations. If we also assume a log-normal distribution, the linear and logarithmic variances are related by:
\begin{align}
\sigma^{2}_{\ln(n_{\rm dust})}(r) &= \ln{\left[ 1 + \sigma^{2}_{n_{\rm dust}}(r)  \right]} \sim \ln{\left[ 1 + \left( \frac{r}{r_{0}} \right)^{-\eta} \right]} \\ 
\sigma_{\log_{10}(n_{\rm dust})} &= \frac{\sigma_{\ln{(n_{\rm dust})}}}{\ln{10}}\sim 0.43\,\sqrt{\ln{\left[ 1 + \left( \frac{r}{r_{0}} \right)^{-\eta} \right]}}
\end{align}
This provides a good approximation to how the PDF of $\delta$ scales versus $h_{\rm min}$; however, from $\xi$ alone we do not have the dependence on $n_{\rm gas}$.

\vspace{-0.5cm}
\section{Discussion}
\label{sec:discussion}

We have shown that aerodynamic particles (e.g.\ dust grains in neutral gas) exhibit large dust-to-gas variations, as well as structure and dynamics {qualitatively} different from the gas, in supersonic, MHD turbulence. 

In some respects, this is similar to the well-studied sub-sonic case in proto-planetary disks. However, a key difference is that in the supersonic case, the gas density exhibits large fluctuations (and the gas-dust velocity contributes to the stopping time). This means that the ``stopping time'' of the dust is no longer constant across the flow, even for a single dust species. We find that this actually {\em enhances} the dynamic range of scales which exhibit dust clustering, in contrast to the case of sub-sonic turbulence, where the dust-to-gas ratio fluctuations tend to be concentrated in a narrow range of scales around the ``resonant'' scale where the eddy turnover time is about equal to the constant stopping time (see references in \S~\ref{sec:intro}). 

We show that fluctuations in the dust-to-gas ratio are approximately log-normal, with two regimes. (1) At low densities $\rho_{\rm gas} < \alpha\,\langle \rho_{\rm gas} \rangle$, the grains de-couple from the gas, so the dust scatters about its mean volume density independent of gas density changes. The nominal dust-to-gas ratios $\rho_{\rm dust}/\rho_{\rm gas}$ in this limit can reach extremely large values, with a power-law tail towards high $\rho_{\rm dust}/\rho_{\rm gas}$. However, the low-density regime is also the limit in which we expect Lorentz forces to begin dominating over drag forces, so the fluctuations may be suppressed. (2) At high densities, the dust and gas are partially coupled. The mean dust density follows the mean gas density; however, there are approximately log-normal fluctuations owing to non-linear grain clustering. Some of this resembles well-studied grain clustering in the sub-sonic limit, since the clustering scales of the dust can be below the sonic scale of the turbulence. But there are additional effects as well, for example, grains can sediment into very thin filaments within shock fronts, similar dynamically to sedimentation under gravitational forces but here the effective acceleration owes to the pressure forces felt by gas and not dust. The magnitude of these fluctuations is large, $\sim 0.3-0.5$\,dex $1\sigma$ dispersion, for grains with size parameter over a wide range $\sizeparam \sim 0.001 - 0.3$, with a maximum around $\sizeparam\sim0.01-0.1$. Much larger grains ($\alpha\gtrsim1$) are never tightly coupled; much smaller grains ($\alpha\lesssim 0.001$) are too well-coupled to gas. The characteristic spatial scales of the grain structures/clustering increase with the grain size (see Eq.~\ref{eqn:Lstream}).

These clustering effects can have many important consequences, which we will explore in future work. For example:

\begin{itemize}

\item{{\bf Dust Formation and Growth:} Because dust is highly-clustered, its growth and evolution, particularly via dust-dust collisions (coagulation or shattering) can be dramatically altered. To lowest order, these effects are manifest in the clumping factor $\langle n_{\rm dust}^{2} \rangle / \langle n_{\rm dust} \rangle^{2}$, which governs the dust-dust interaction rate and reaches values $\sim 50-100$. The effects of grain clustering on growth have been extensively studied in proto-planetary disks; however they are not well-understood in molecular clouds. At the very least, the large clumping factors imply order-of-magnitude faster evolution of large grains in neutral clouds compared to what is usually assumed. Other effects, for example the non-uniform and size-dependent velocity dispersions of grains, may substantially alter both collision rates and the outcomes of those collisions (sticking vs.\ shattering, for example). These effects, in turn, may dramatically influence the size distribution of dust.}

\item{{\bf Extinction Mapping and Dust Emission:} Visually, it is obvious that the dust and gas are not necessarily co-located. In probes of extinction and dust emission, this may be directly visible; however we caution that the predictions here correspond to {\em large} dust grains. These do not dominate extinction. Rather, one would have to use diagnostics specifically sensitive to large grains (for example, sub-mm observations). Moreover, there is a finite scale which must be resolved in order to see the dust-to-gas fluctuations. On larger scales compared to the critical scale for dust clustering, they will be smoothed out, and one will simply trace the mean dust-to-gas ratio. But such fluctuations, on scales similar to the critical scale predicted, have been observed in many nearby clouds and some centers of nearby galaxies including e.g.\ Taurus \citep{padoan:dust.fluct.taurus.vs.sims,flagey:2009.taurus.large.small.to.large.dust.abundance.variations,pineda:2010.taurus.large.extinction.variations}, NGC 1266 \citep{pellegrini:2013.ngc.1266.shocked.molecules,nyland:2013.radio.core.ngc1266}, Orion \citep{abergel:2002.size.segregation.effects.seen.in.orion.small.dust.abundances}, the Ursa Major cirrus \citep{miville-deschenes:2002.large.fluct.in.small.grain.abundances}, IC 5146, CGCG525-46, IR04139+2737, and G0858+723 \citep{thoraval:1997.sub.0pt04pc.no.cloud.extinction.fluct.but.are.on.larger.scales,thoraval:1999.small.scale.dust.to.gas.density.fluctuations}. The absolute scales where the fluctuations are observed range from $\sim 0.003 - 10$\,pc, but in each case the critical scale and magnitude of fluctuations appears to agree with the simple scalings expected for turbulent concentration, given the different cloud densities and grain sizes probed (for detailed comparisons, see \citealt{padoan:dust.fluct.taurus.vs.sims} and \citealt{hopkins:totally.metal.stars}). Thus, great care is needed, especially as observations push to higher resolution. Both dust and gas have a filamentary morphology, but dust filaments may, in fact, be much narrower than gas filaments (which are characteristically of order the sonic length); in rare cases, dust filaments can exist where there is no gas filament at all (owing to dust concentration by gas vorticity). Very large dust grains ($\gg1\,\mu$m), on the other hand, may be more uniformly distributed than gas throughout clouds. This may resolve several long-standing puzzles regarding apparently different extinction measurements that have alternatively been attributed to different dust chemistry in different regions.}

\item{{\bf Cooling Physics \&\ Star Formation:} In dense regions of clouds or galactic nuclei, gas cooling or heating can be regulated by collisions with dust, with the relevant rate proportional to the local dust-to-gas ratio. When this dominates, we therefore predict that there may be order-of-magnitude variations in the cooling physics of some regions. In metal-poor galaxies, regions which are relatively over-abundant in dust may be preferentially able to form stars, since low-mass star formation may be difficult without sufficient dust present to act as a coolant.}

\item{{\bf Stellar Abundances:} As proposed in \citet{hopkins:totally.metal.stars}, large fluctuations in the local dust-to-gas ratio should translate to interesting variations in stellar abundances, even for stars formed in the same cluster. Large dust grains contain most of the dust mass (about $\sim1/2$ the total metal mass), and they are the ones for which these fluctuations are important. Even smoothing on relatively large scales, corresponding to $\gtrsim 0.1\,$pc (the size of large protostellar cores), we predict significant fluctuations if grains have the appropriate sizes. Specifically, assuming $L_{\rm box}\sim10\,$pc and $\Sigma_{\rm GMC}\sim300\,M_{\sun}\,{\rm pc}^{-2}$, and that $1/3$ the metals are in grains with sizes $\sim 0.1\,\mu\,$m, we predict an approximately $\approx 0.05-0.1$\,dex $1\sigma$ dispersion in the total metallicity of the dense regions (owing to dust-to-gas fluctuations); this is small and well within the dispersion observed for nearby clusters \citep{casagrande:2011.geneva.survey.update.metal.rich.wing.of.local.stars,duran:2013.local.age.metallicity.relation}. More interestingly, though, because it is log-normal, the distribution has a long tail, and one dense star-forming region per million could have a total metallicity enhancement of a factor $\sim 20-50$!}

\end{itemize}

Studying these in more detail requires additional simulations with the relevant physics included, which makes our calculations no longer scale-free. However, it is straightforward to extend our models and follow these additional processes. Moreover, applying these simulations to a specific scale and situation allows for the inclusion of additional, non scale-free physics which we have ignored in this first study (such as Lorentz forces and grain collisions).

\vspace{-0.7cm}
\acknowledgments 
We thank Jim Stone, Paolo Padoan, Jessie Christiansen, Charlie Conroy, Evan Kirby, and Paul Torrey for helpful discussions and contributions motivating this work. We also thank the anonymous referee for valuable suggestions. Support for PFH was provided by the Gordon and Betty Moore Foundation through Grant \#776 to the Caltech Moore Center for Theoretical Cosmology and Physics, an Alfred P. Sloan Research Fellowship, NASA ATP Grant NNX14AH35G, and NSF Collaborative Research Grant \#1411920. Numerical calculations were run on the Caltech compute cluster ``Zwicky'' (NSF MRI award \#PHY-0960291) and allocation TG-AST130039 granted by the Extreme Science and Engineering Discovery Environment (XSEDE) supported by the NSF.\\ 

\bibliography{/Users/phopkins/Dropbox/Public/ms}

\begin{appendix}

\section{Distributions Per Unit Volume and Per Unit Gas Mass}
\label{sec:alt.pdfs}

\begin{figure*}
    \plotsidesize{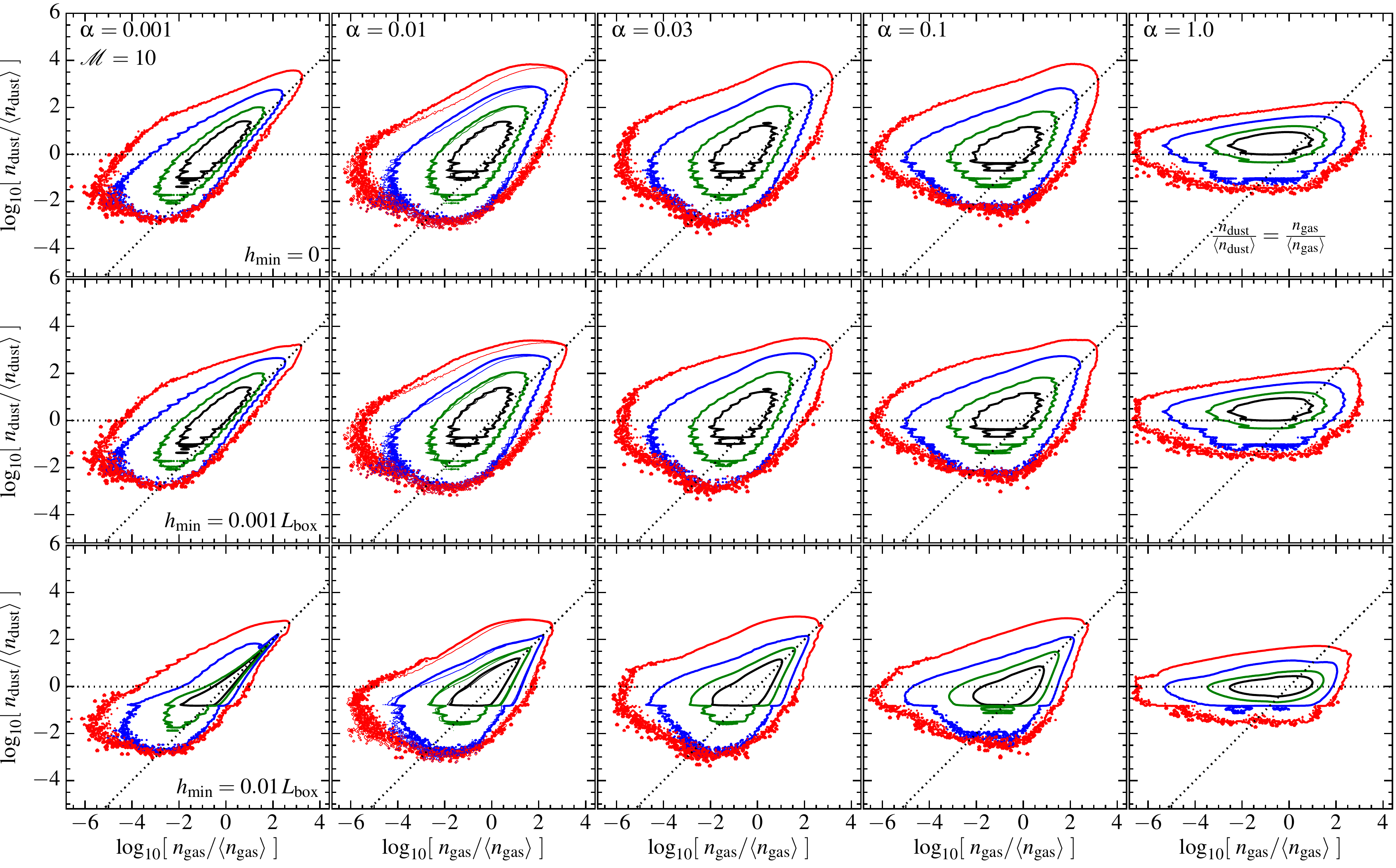}{0.9}
    \vspace{-0.25cm}
    \caption{Distribution of dust and gas densities in our $\mathcal{M}\sim10$ simulations, as Fig.~\ref{fig:ndust.ngas.m10} in the text. The only difference is that here, we measure the probability distribution function around random points in space within the box (i.e.\ measure the volume-weighted PDF $dP_{\rm vol}$), instead of the distribution around random dust particles ($dP_{\rm dust}$). For $h_{\rm min}=0$, the two are trivially related by $dP_{\rm vol} \propto n_{\rm dust}^{-1}\,dP_{\rm dust}$. The contours shift to lower $n_{\rm dust}$ and $n_{\rm gas}$ as these have larger volume-filling factors, and the peak of the volume-averaged ($h_{\rm min}>0$) probability density shifts closer to $n_{\rm gas}=\langle n_{\rm gas} \rangle$, $n_{\rm dust}=\langle n_{\rm dust} \rangle$, as it must, but the qualitative behavior and scatter in $n_{\rm dust}$ at fixed $n_{\rm gas}$ is similar in all cases. 
    \vspacerpostplot 
    \label{fig:ndust.ngas.m10.pvol}}
\end{figure*}

\begin{figure*}
    \plotsidesize{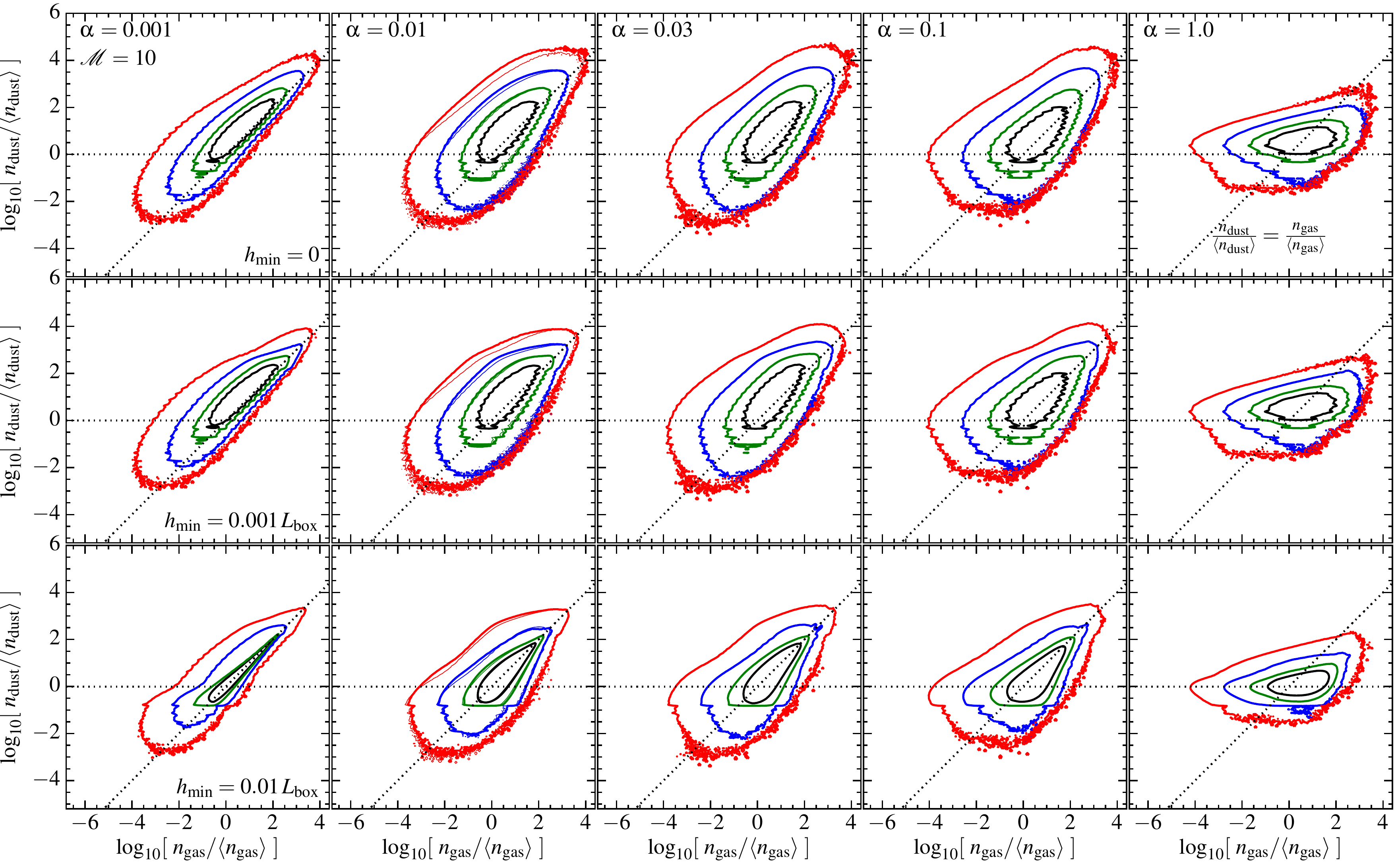}{0.9}
    \vspace{-0.25cm}
    \caption{Distribution of dust and gas densities in our $\mathcal{M}\sim10$ simulations, as Fig.~\ref{fig:ndust.ngas.m10} in the text. As Fig.~\ref{fig:ndust.ngas.m10.pvol}, the difference here is that we measure the PDF around random gas elements (i.e.\ the gas-mass weighted PDF $dP_{\rm gas}$) instead of around random dust elements ($dP_{\rm dust}$) or random volume elements ($dP_{\rm vol}$). For $h_{\rm min}=0$ these are related by $dP_{\rm gas} \propto n_{\rm gas}\,dP_{\rm vol} \propto (n_{\rm gas}/n_{\rm dust})\,dP_{\rm dust}$. Overall, the contours shift noticeably towards the mean dust-to-gas ratio (especially at low densities), i.e.\ most of the gas sees a ratio closer to the mean, compared to what most of the {\em dust} sees, because the dust is more highly-clustered than the gas. The scatter about this mean at high $n_{\rm gas}$, however, is similar in $dP_{\rm gas}$ and $dP_{\rm dust}$. 
    \vspacerpostplot 
    \label{fig:ndust.ngas.m10.pgas}}
\end{figure*}

In the text, we have (unless otherwise specified) shown the probability distribution function (PDF) of field properties such as $n_{\rm dust}$ and $n_{\rm gas}$ {\em around a random dust grain}. This is the dust-mass weighted PDF $dP_{\rm dust}$:
\begin{align}
dP_{\rm dust} = {\rm Probability(around\ random\ grain)}
\end{align}
Because, within an infinitesimally small differential volume $d^{3}{\bf x} = dV$, the dust properties are sampled $N=n_{\rm dust}\,dV$ times, it is trivial to show that this is related to the PDF {\em around a random point in space} ${\bf x}$, i.e.\ the volume-weighted PDF $dP_{\rm vol}$, by:
\begin{align}
dP_{\rm vol} \propto n_{\rm dust}^{-1}\,dP_{\rm dust}
\end{align}
Note that this trivial conversion is only strictly true if we measure the point-like density $n_{\rm dust}$ -- if we instead smooth the properties on some finite averaging scale $h_{\rm min}>0$, this is only approximate. However, we can, of course, rigorously calculate $dP_{\rm Vol}$ by following the same measurement procedure described in the text, beginning from randomly-selected, uniformly-distributed points in space, as opposed to the locations of dust grains.
Finally, it is likewise trivial to show that the probability {\em around a random gas element}, i.e.\ the Lagrangian or gas-mass weighted PDF $dP_{\rm gas}$, is given by:
\begin{align}
dP_{\rm gas} \propto n_{\rm gas}\,dP_{\rm vol}
\end{align}

Figs.~\ref{fig:ndust.ngas.m10.pvol}-\ref{fig:ndust.ngas.m10.pgas} show the full bivariate distribution of $n_{\rm dust}$ and $n_{\rm gas}$ as Fig.~\ref{fig:ndust.ngas.m10} in the text. However, Fig.~\ref{fig:ndust.ngas.m10.pvol} shows the volume-weighted probability $P_{\rm vol}$ and Fig.~\ref{fig:ndust.ngas.m10.pgas} shows the gas-mass weighted probability $P_{\rm gas}$. These are calculated correctly for $h_{\rm min}>0$ but are very close to the approximate values given by the simple relations above.

First consider $dP_{\rm vol}$. As expected, the distribution shifts to lower $n_{\rm dust}$ and $n_{\rm gas}$ as these regions have larger volume-filling factors. Similarly the peak of the probability density shifts closer to $n_{\rm gas}=\langle n_{\rm gas} \rangle$, $n_{\rm dust}=\langle n_{\rm dust} \rangle$, and rapidly moves onto this point as we increase the volume-averaging scale $h_{\rm min}$ (since the distribution function for even infinitely-clustered dust must converge to a delta function around this point as $h_{\rm min}\rightarrow L_{\rm box}$). Modulo the mildly reduced scatter towards high $n_{\rm dust}$, however, the qualitative behavior of the distribution functions is similar to $dP_{\rm dust}$ in all cases. 

Next consider $dP_{\rm gas}$. Interestingly, in this case, the distribution shifts significantly towards the mean dust-to-gas ratio, especially at low densities. This is partly because the low-$n_{\rm gas}$ regions (where the dust de-couples) are down-weighted in the distribution. It cannot entirely be explained by this effect however -- even at intermediate/high densities, most of the gas has a local dust-to-gas ratio closer to the mean compared to most of the dust. As noted in the main text, this is expected because the dust is more highly-clustered than the gas. But once again, the high-$n_{\rm gas}$ behavior and scatter is still similar to $dP_{\rm dust}$.

A simple model explains these results. In super-sonic turbulence, the gas density is approximately log-normal, with $dP_{\rm vol} \propto \exp{[-(\ln{n_{\rm gas}}-S_{\rm gas}/2)^{2}/(2\,S_{\rm gas})]}$ where $S_{\rm gas}\approx \ln{[1+b^{2}\,\mathcal{M}^{2}]}$ is the variance ($b\sim1/2$, depending on details of the turbulence). Since the dust does not modify the gas in our runs, the bivariate distribution should reflect this for the gas, with $P(n_{\rm dust}\,|\,n_{\rm gas})$ also approximately log-normal, as shown in Fig.~\ref{fig:dustgas.pvsweights}. But for a log-normal distribution $dP(\ln{x})$, it is trivial to show that the distribution $dP_{\rm new}\propto x\,dP$ is {\em also} log-normal, with the same variance but a shifted mean. Therefore, in this case, we expect the bivariate distributions to be approximately given by:
\begin{align}
\frac{dP_{\rm vol}}{d\ln{n_{\rm gas}}\,d\ln{n_{\rm dust}}} &=
\frac{1}{2\pi\sqrt{S_{\rm gas}\,S_{\rm dust}}}\,
\exp{\left[ -\frac{\Delta_{\rm gas}^{2}}{2\,S_{\rm gas}} - \frac{\tilde{\Delta}_{\rm dust}^{2}}{2\,S_{\rm dust}} \right]}\\
\frac{dP_{\rm gas}}{d\ln{n_{\rm gas}}\,d\ln{n_{\rm dust}}} &=
\frac{1}{2\pi\sqrt{S_{\rm gas}\,S_{\rm dust}}}\,
\exp{\left[ -\frac{\tilde{\Delta}_{\rm gas}^{2}}{2\,S_{\rm gas}} - \frac{\tilde{\Delta}_{\rm dust}^{2}}{2\,S_{\rm dust}} \right]}\\
\frac{dP_{\rm dust}}{d\ln{n_{\rm gas}}\,d\ln{n_{\rm dust}}} &=
\frac{1}{2\pi\sqrt{S_{\rm gas}\,S_{\rm dust}}}\,
\exp{\left[ -\frac{\Delta_{\rm gas}^{2}}{2\,S_{\rm gas}} - \frac{{\Delta}_{\rm dust}^{2}}{2\,S_{\rm dust}} \right]}\\
 \Delta_{\rm gas} &\equiv \ln{\left(\frac{n_{\rm gas}}{\langle n_{\rm gas}\rangle}\right)} + \frac{S_{\rm gas}}{2}\\
 \tilde{\Delta}_{\rm gas} &\equiv {\Delta}_{\rm gas}-S_{\rm gas} = \ln{\left(\frac{n_{\rm gas}}{\langle n_{\rm gas}\rangle}\right)} - \frac{S_{\rm gas}}{2}\\
 {\Delta}_{\rm dust} & \equiv \ln{(\delta)} - \langle \ln{\delta(n_{\rm gas})} \rangle_{\rm dust} \\
 \tilde{\Delta}_{\rm dust} &\equiv \Delta_{\rm dust} + S_{\rm dust}  = \ln{(\delta)} - \delta_{0} 
\end{align}
where $S_{\rm gas}$ is constant but $S_{\rm dust}=S_{\rm dust}(n_{\rm gas})$ and $\delta_{0}=\delta_{0}(n_{\rm gas})$ are functions of $n_{\rm gas}$, and 
\begin{align}
\delta &\equiv \frac{n_{\rm dust}/n_{\rm gas}}{\langle n_{\rm dust} \rangle / \langle n_{\rm gas} \rangle}\\
\delta_{0} &\equiv \langle \ln \delta(n_{\rm gas})\rangle_{\rm vol} = \langle \ln \delta(n_{\rm gas})\rangle_{\rm dust}-S_{\rm dust}
\end{align}
It is easy to verify these obey $dP_{\rm gas} \propto n_{\rm gas}\,dP_{\rm vol} \propto (n_{\rm gas}/n_{\rm dust})\,dP_{\rm dust}$. Trivially, therefore, if the distributions are lognormal in $\delta$ and $n_{\rm gas}$, the logarithmic scatter is identical regardless of how we weight the distributions, and the mean values simply shift. 

The values of $S_{\rm dust}$ and $\delta_{0}$ can be read off of Fig.~\ref{fig:variance}, noting $S_{\rm dust} = [\sigma_{\log{10}}(\delta)\,\ln{10}]^{2} \sim 0.5-2$, and that the plotted $\langle \log_{10} \delta(n_{\rm gas})\rangle_{\rm dust}=(\delta_{0}+S_{\rm dust})/\ln{(10)}$. If $\delta_{0}$ and $S_{\rm dust}$ are constant (approximately true in the high-density limit), then the constraint that the PDF integrates correctly gives $\delta_{0} = -S_{\rm dust}/2$; if we instead assume $S_{\rm dust}$ is constant but $\delta_{0} = A - \ln{(n_{\rm gas}/\langle n_{\rm gas}\rangle)}$ (i.e.\ $n_{\rm dust}\sim\,$constant, approximately true in the low-density limit), we have $A = -S_{\rm dust}/2$. These give good approximations in both limits to the results in Fig.~\ref{fig:variance}.

\section{Resolution Study}
\label{sec:resolution}

\begin{figure}
 \begin{tabular}{c}
  \vspace{0.1cm}
  \includegraphics[width=0.9\columnwidth]{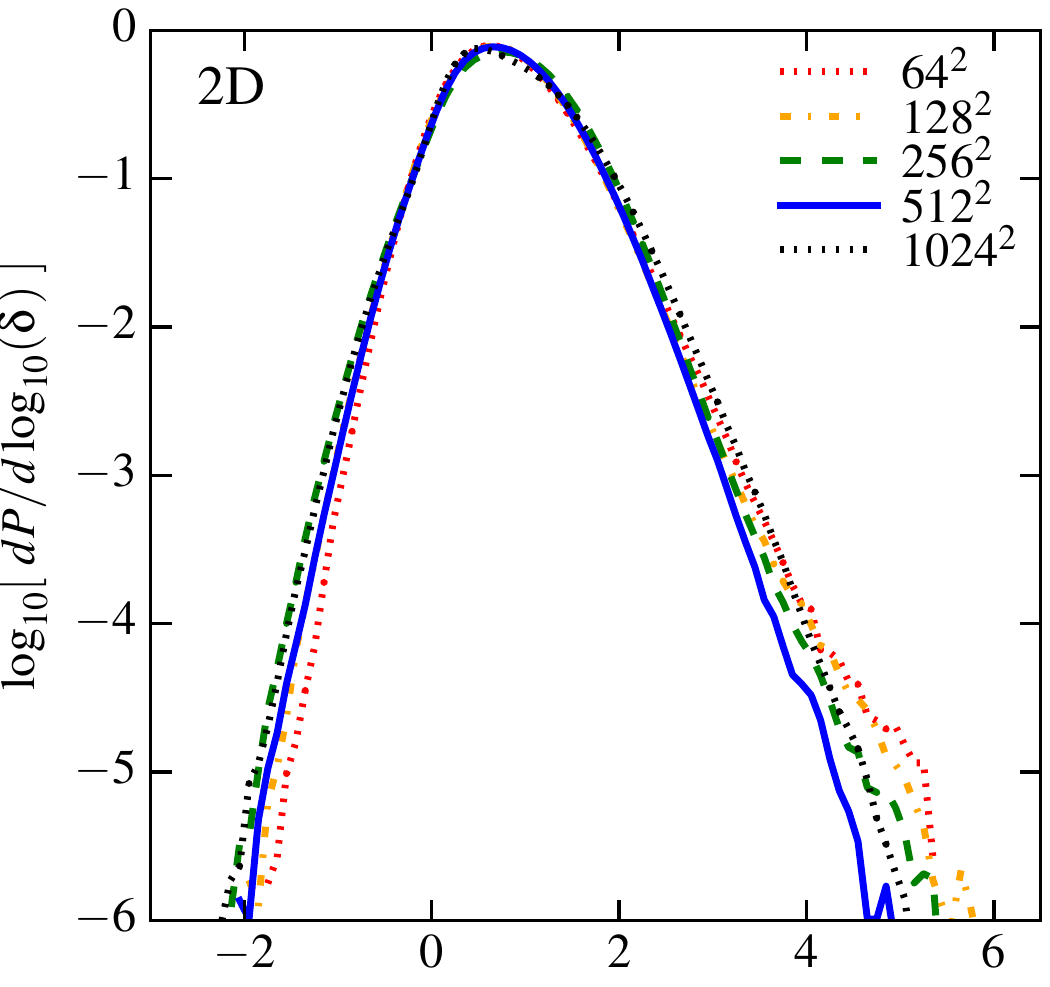} \\
  \includegraphics[width=0.9\columnwidth]{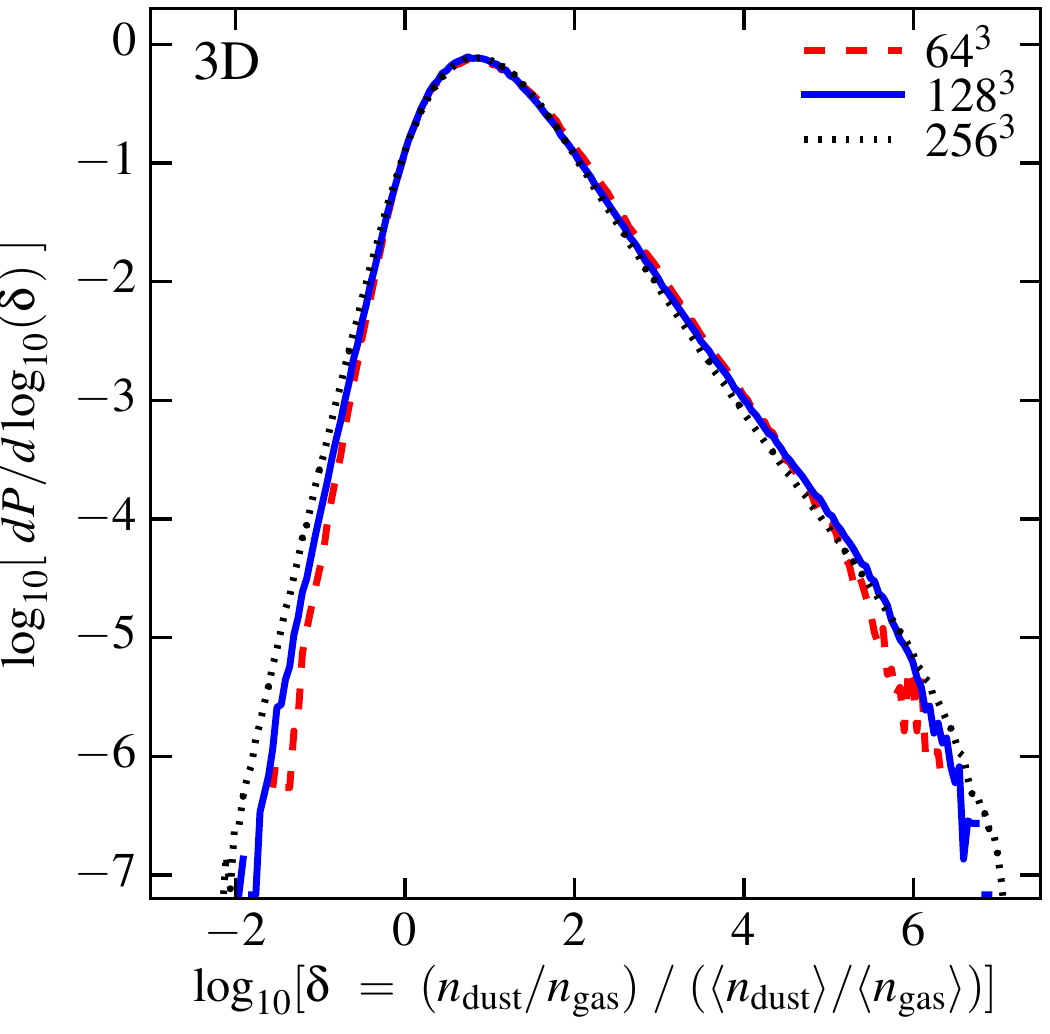} \\
 \end{tabular}
    \vspace{-0.25cm}
    \caption{{\em Top:} Distribution of dust-to-gas ratios (as Fig.~\ref{fig:dustgas.pdf}), in a 2D resolution study with $\mathcal{M}_{2D}\sim5$ and $\sizeparam=0.01$. 
    {\em Bottom:} Same, in a 3D study with $\mathcal{M}\sim10$ and $\sizeparam=0.03$. 
    Owing to the Lagrangian nature of our code, and to the fact that the turbulence is super-sonic (and so structures are driven by relatively easily-captured shocks and rarefactions), the convergence is remarkably good. Even $64^{D}$ runs appear well-converged in the core of the PDF; by $256^{D}$ the results agree well with our $1024^{D}$ simulations (in 2D). We expect our conclusions in the text are robust to resolution effects. 
        \vspacerpostplot 
    \label{fig:resolution}}
\end{figure}

Obviously it is important to test that our simulations are numerically converged. Because the resolution we can achieve is more limited in 3D, we consider a resolution study first in 2D reaching much higher resolution than our standard runs in the text. 

Fig.~\ref{fig:resolution} plots the time-averaged PDF of the dust-to-gas ratio, in the style of Fig.~\ref{fig:dustgas.pdf} in the text, for 2D simulations with Mach number $\mathcal{M}_{2D}\sim5$ and $\sizeparam=0.01$. We consider resolutions $64^{2}-1024^{2}$. As expected, the tails of the PDF are better sampled at higher resolution -- this follows from simple counting statistics. Remarkably, the core of the PDF appears reasonably well-converged at just $\sim64^{2}$ resolution; by $\sim 256^{2}$ the ``wings'' agree well down to part-per-million amplitudes (there is a small deviation in the $512^{2}$ run, such that the $1024^{2}$ run actually agrees slightly better with the $256^{2}$ run; this appears to depend on how the turbulent driving routines depend on resolution). This justifies our choice in the text of $256^{3}$ resolution. 

Interestingly, the convergence here is much faster than seen in some previous studies \citep[compare e.g.][]{bai:2010.grain.streaming.sims.test}. This owes in part to the Lagrangian nature of our method, which is able, in principle, to capture arbitrarily large fluctuations in density (so long as they involve equal to or larger than some fixed mass scale) at low ``resolution'' (i.e.\ there is no fixed spatial resolution). It also owes to the super-sonic nature of the turbulence here, where much of the dynamics is driven by shocks and rarefactions (relatively ``easily'' captured in these methods), as opposed to the streaming instability or details of the vorticity field of small turbulent eddies (the dominant effects in the highly-subsonic limit). 

We have also considered a limited study in 3D, taking one of our standard runs and re-running at lower resolution. Even at $64^{3}$, our qualitative conclusions are essentially identical (although the extremes of the distribution functions are sampled relatively poorly).

\vspace{-0.5cm}
\section{Numerical Effects of the Dust Drag Algorithm \&\ Poisson Errors}
\label{sec:dust.algorithm}

\begin{figure}
 \begin{tabular}{r}
  \vspace{0.1cm}
  \includegraphics[width=0.865\columnwidth]{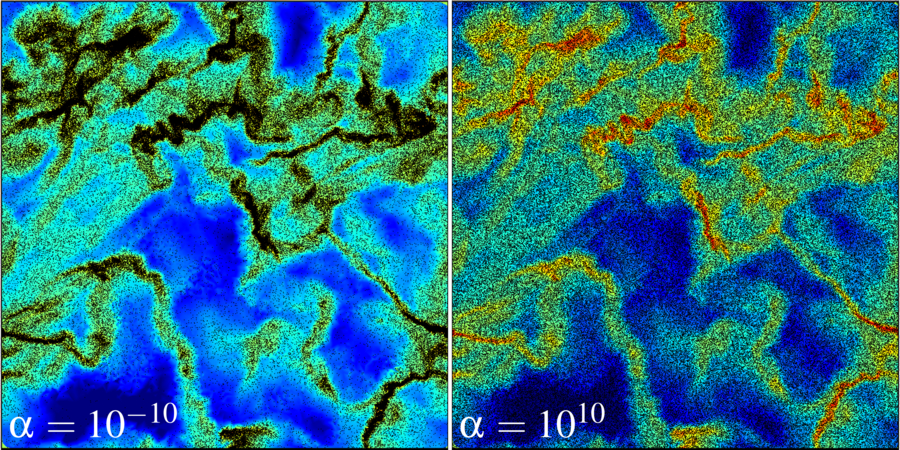} \\
  \includegraphics[width=0.97\columnwidth]{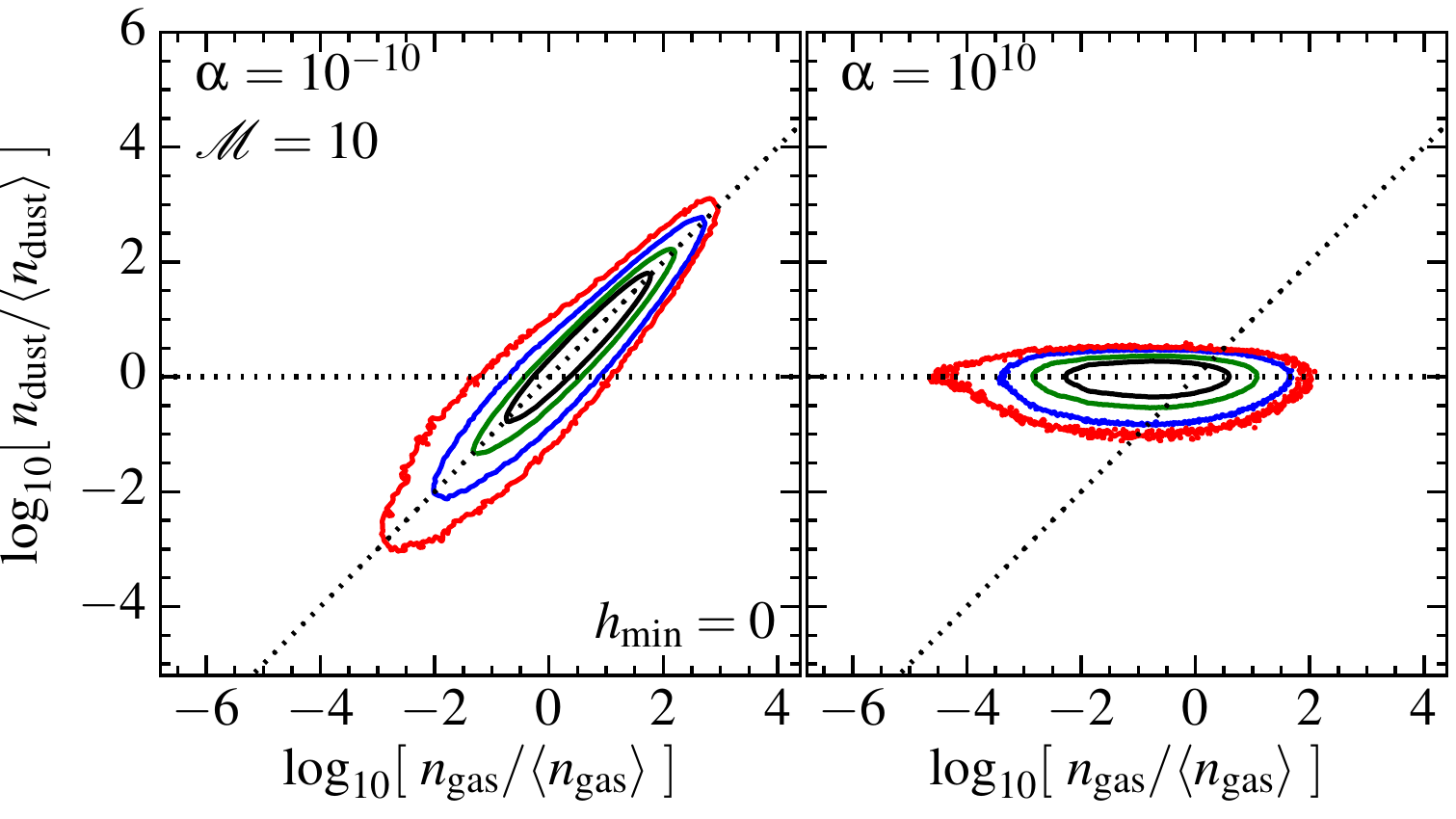} \\
  \includegraphics[width=0.97\columnwidth]{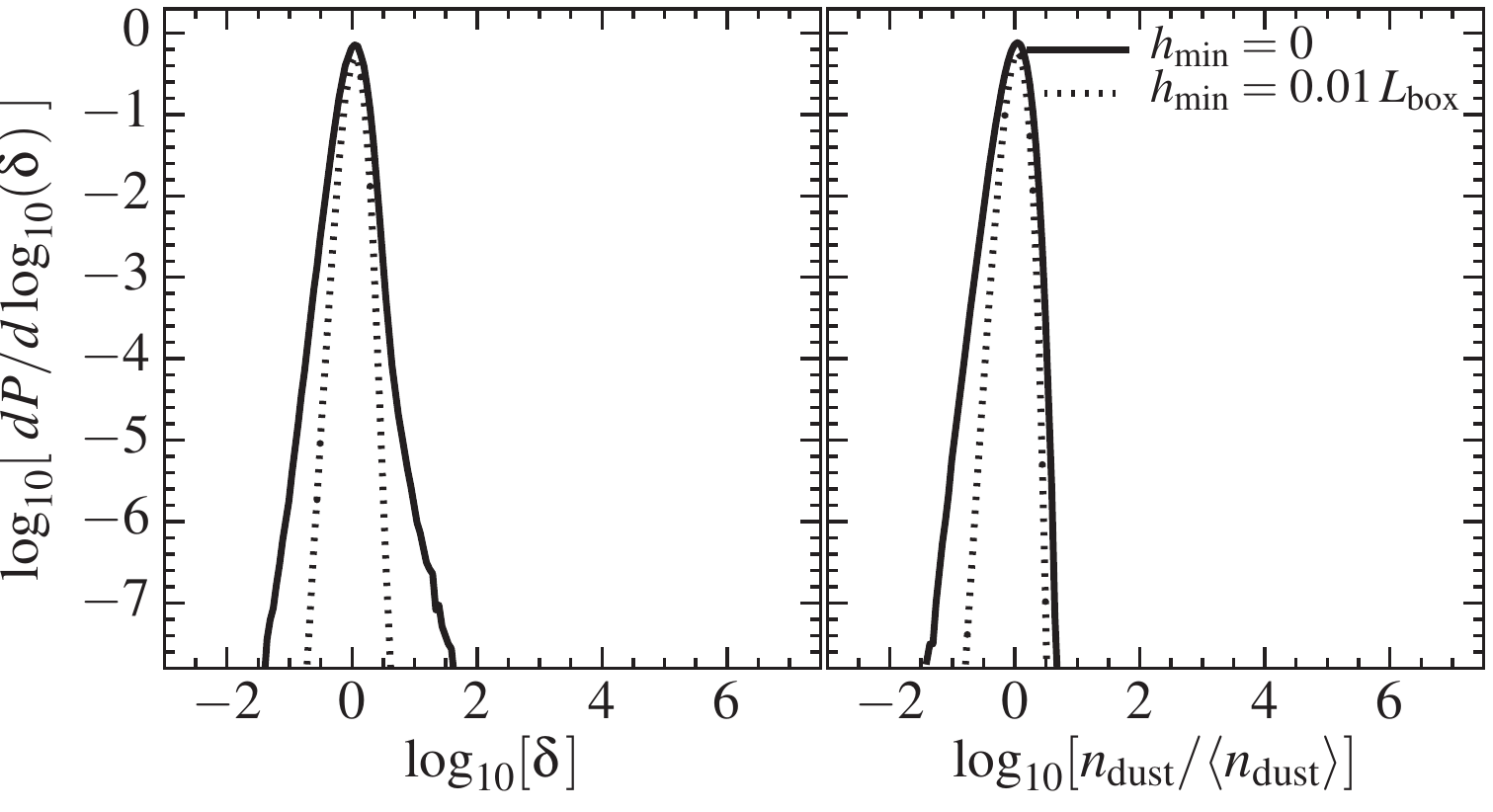} 
 \end{tabular}
    \vspace{-0.25cm}
    \caption{{\em Top:} Images of the dust and gas density as Fig.~\ref{fig:demo.vs.alpha}, in a $\mathcal{M}\sim10$ simulation with almost perfectly-coupled $\alpha=10^{-10}$ ({\em left}) and almost perfectly-uncoupled $\alpha=10^{10}$ ({\em right}). 
    {\em Middle:} Bivariate dust and gas distribution as Fig.~\ref{fig:ndust.ngas.m10} for both cases. We show $h_{\rm min}=0$; the scatter decreases for larger $h_{\rm min}$.  
    {\em Bottom:} Histogram of the dust-to-gas ratio $\delta$ ({\em left}) and dust density $n_{\rm dust}$ ({\em right}), as Fig.~\ref{fig:dustgas.pdf}. 
    In the perfectly-coupled case, dust should track gas exactly ($\delta=1$), in the un-coupled case, dust should remain at the mean density ($n_{\rm dust}=\langle n_{\rm dust} \rangle$). Poisson sampling from our finite particle number (laid down randomly in the initial conditions) leads to some scatter. For the strongly-coupled case, these errors are enhanced by small numerical differences between the algorithms used to update the gas and dust velocities. However, the distributions do not show any systematic deviation from the expected behavior. Their widths ($\sigma\sim 0.05-0.07$\,dex) are much smaller than any $\alpha\sim0.001-1$ case we consider in the text, so the errors are not significant in our study. 
        \vspacerpostplot 
    \label{fig:test.alpha.limits}}
\end{figure}

Although the scheme used here to integrate the trajectories of dust particles is standard and relatively well-tested \citep[similar to][]{carballido:2008.grain.streaming.instab.sims,hogan:1999.turb.concentration.sims,johansen:2007.streaming.instab.sims,johansen:2009.particle.clumping.metallicity.dependence,bai:2010.grain.streaming.sims.test,pan:2011.grain.clustering.midstokes.sims}, there are known sources of numerical error. 

The advantage of a Lagrangian ``super-particle'' approach is that, in the limit where the grains are decoupled from the gas ($\alpha\rightarrow\infty$), their dynamics (free-streaming) are perfectly recovered, and the only source of error in the density field is Poisson noise (from our finite particle number). This is not true in ``two-fluid'' approximations, for example, which cannot account for the full velocity distribution function of grains at a single location. 

In the opposite limit of perfect coupling ($\alpha\rightarrow 0$), the grains should perfectly trace the gas (as tracer particles), up to Poisson noise in the initial tracer field. However, our methods introduce an additional error: when $\alpha\rightarrow0$, the algorithm used to update the particle velocities and positions (interpolating to the particle position) does not, numerically, perfectly match the Godunov-type update to the gas particle velocities (involving the solution of a Riemann problem). In a sufficiently smooth flow, these should be identical, but given numerical noise or physical discontinuities, they can differ \citep[for detailed analysis of these errors, see][]{genel:tracer.particle.method}. 

We therefore test both limits here. We take our standard $\mathcal{M}=10$ simulation and re-run with $\alpha=10^{10}$ (effectively infinite) and $\alpha=10^{-10}$ (effectively zero). In Fig.~\ref{fig:test.alpha.limits}, we plot the resulting images, bivariate density distributions, and time-averaged PDF of the dust density (for $\alpha=10^{10}$) and dust-to-gas ratio (for $\alpha=10^{-10}$). We take $h_{\rm min}=0$, since the errors of interest rapidly become smaller as the averaging scale becomes larger. For $\alpha=10^{10}$, we confirm that the scatter in dust density is what we expect from Poisson statistics (with smaller residual errors owing to our post-processing kernel density estimator). For $\alpha=10^{-10}$, we find the dust traces gas at all densities, with a comparable scatter to the Poisson case. 

In both cases, the scatter in the core of the distribution is $< 0.1$\,dex; much smaller than we see in any of our simulations with $0.001\lesssim \alpha\lesssim 1$. Moreover, the tails of the distribution are dramatically suppressed -- these are many orders of magnitude smaller than we see in the text. And in both cases, the mean dust density behaves as it should and we see no unphysical features (only noise).\footnote{We note that the errors in the ``tracer particle limit'' $\alpha=10^{-10}$ here are significantly smaller than those shown in \citet{genel:tracer.particle.method}. This owes to several effects: we use a smooth (as opposed to discontinuous) interpolation for the velocity field, our method is fully Lagrangian (there are no inter-particle mass fluxes to enhance discrepancies in advection), we use the exact solution over the timestep to update particle velocities (as opposed to only the instantaneous acceleration), and we synchronize the time updates between gas and dust (and use a stricter dust timestep criterion). As discussed therein, these all reduce (although do not completely eliminate) the relevant errors.}

We conclude that these sources of error are not significant for the $\alpha$ values in the text. Based crudely on the scaling of the variance in Fig.~\ref{fig:variance}, we estimate that Poisson noise and/or integration errors would, at our current resolution, become significant compared to physical effects at $\alpha \ll  10^{-4}$ or $\alpha \gg 100$, necessitating higher-resolution studies.

\end{appendix}

\end{document}